\documentclass[%
reprint,
superscriptaddress,
amsmath,amssymb,
aps,pra
twocolumn,
showpacs,
]{revtex4-1}

\usepackage{tikz}
\usepackage{graphicx}
\usepackage{dcolumn}
\usepackage{bm}
\usepackage{braket}
\usepackage{hyperref}
\usepackage[caption=false]{subfig}
\usepackage{amsmath}
\usepackage{graphicx}
\usepackage{fancyhdr}
\usepackage{float}
\usepackage[separate-uncertainty=true]{siunitx}
\usepackage{stackengine}

\DeclareSIUnit{\dBm}{\deci\belmilliwatt}
\DeclareSIUnit{\gammas}{$\Gamma$}
\usepackage{xcolor}
\usepackage{xspace}
\usepackage{comment}
\usepackage{eucal}
\usepackage{multirow}
\usepackage[utf8]{inputenc}
\usepackage{pgfplots}
\pgfplotsset{compat=1.14}

\newcommand{\mf}{m_F\xspace}

\newcommand{\DOne}{D\textsubscript{1}\xspace}

\newcommand{\Rb}{\textsuperscript{87}Rb\xspace}
\newcommand{\K}{\textsuperscript{39}K\xspace}

\DeclareSIUnit{\gauss}{G}

\begin{document}

\newcommand{\update}[1]{\textbf{\color{red}UPDATE:~#1}\xspace}
\newcommand{\strike}[1]{\color{red}\sout{#1}\xspace}
\newcommand{\new}[1]{{\color{green}#1}\xspace}
\sisetup{detect-weight=true, detect-family=true}

\title{
A high-flux source system for matter-wave interferometry exploiting tunable interactions
}
\author{A.~Herbst}
\affiliation{Leibniz Universit\"at Hannover, Institut f\"ur Quantenoptik,\\ Welfengarten 1, 30167 Hannover, Germany}
\author{T.~Estrampes}
\affiliation{Leibniz Universit\"at Hannover, Institut f\"ur Quantenoptik,\\ Welfengarten 1, 30167 Hannover, Germany}
\affiliation{Universit\'e Paris-Saclay, CNRS, Institut des Sciences Mol\'eculaires d'Orsay, 91405 Orsay, France}
\author{H.~Albers}
\affiliation{Leibniz Universit\"at Hannover, Institut f\"ur Quantenoptik,\\ Welfengarten 1, 30167 Hannover, Germany}
\author{V.~Vollenkemper}
\affiliation{Leibniz Universit\"at Hannover, Institut f\"ur Quantenoptik,\\ Welfengarten 1, 30167 Hannover, Germany}
\author{K.~Stolzenberg}
\affiliation{Leibniz Universit\"at Hannover, Institut f\"ur Quantenoptik,\\ Welfengarten 1, 30167 Hannover, Germany}
\author{S.~Bode}
\affiliation{Leibniz Universit\"at Hannover, Institut f\"ur Quantenoptik,\\ Welfengarten 1, 30167 Hannover, Germany}
\author{E.~Charron}
\affiliation{Universit\'e Paris-Saclay, CNRS, Institut des Sciences Mol\'eculaires d'Orsay, 91405 Orsay, France}
\author{E.~M.~Rasel}
\affiliation{Leibniz Universit\"at Hannover, Institut f\"ur Quantenoptik,\\ Welfengarten 1, 30167 Hannover, Germany}
\author{N.~Gaaloul}
\affiliation{Leibniz Universit\"at Hannover, Institut f\"ur Quantenoptik,\\ Welfengarten 1, 30167 Hannover, Germany}
\author{D.~Schlippert}\email[Electronic mail: ]{schlippert@iqo.uni-hannover.de}
\affiliation{Leibniz Universit\"at Hannover, Institut f\"ur Quantenoptik,\\ Welfengarten 1, 30167 Hannover, Germany}

\date{\today}

\begin{abstract}
Atom interferometers allow determining inertial effects to high accuracy. 
Quantum-projection noise as well as systematic effects impose demands on large atomic flux as well as ultra-low expansion rates.
Here we report on a high-flux source of ultra-cold atoms with free expansion rates near the Heisenberg limit directly upon release from the trap. 
Our results are achieved in a time-averaged optical dipole trap and enabled through dynamic tuning of the atomic scattering length across two orders of magnitude interaction strength via magnetic Feshbach resonances. 
We demonstrate BECs with more than \num{6e4} particles after evaporative cooling for 170 ms and their subsequent release with a minimal expansion energy of 4.5 nK in one direction.   
Based on our results we estimate the performance of an atom interferometer and compare our source system to a high performance chip-trap, as readily available for ultra-precise measurements in micro-gravity environments.
\end{abstract}
\maketitle
\section{Introduction}
Quantum sensors based on atom interferometry~\cite{Kasevich91PRL, Riehle91PRL, Kasevich92APB, Cronin09RMP} allow for the  absolute determination of inertial effects to great accuracy~\cite{Gustavson97PRL, Canuel06PRL, Dickerson13PRL, Dutta16PRL, Savoie2018SciAdv}.
As such, they hold enormous potential across a broad spectrum of research areas, encompassing earth observation, environmental monitoring, navigation, and resource exploration.
In fundamental physics they have been successfully used to test the weak equivalence principle~\cite{Schlippert14PRL, Tarallo2014PRL, Albers2020EPJD, Asenbaum2020PRL}, challenge the fundamental assumptions of quantum mechanics~\cite{Bassi2013RMP, Kovachy15Nature, Carlesso2022NatPhys, Schrinski23PRA}, and to determine fundamental constants~\cite{Rosi14Nature, Parker2018Science, Morel2020Nature}.
Recent proposals now aim for the search for dark matter~\cite{ElNeaj2020EPJQ, Du2022PRD, Badurina2023PRD} and the detection of gravitational waves in frequency bands complementary to those accessible using laser interferometers~\cite{Hogan2011GRG, Canuel2018SciRep, Zhan2019quq, schubert_scalable_2019, Canuel2020CQG, Badurina2020JCA}. 

In order to reach the required sensitivity levels, a high atomic flux in combination with an extended pulse separation time is necessary. 
To enhance the latter the use of Bose-Einstein condensates (BECs)~\cite{Anderson95Science,Davis95PRL} presents a viable approach. 
In comparison to thermal ensembles, BECs allow for a superior control of systematic errors, yield higher signal-to-noise ratios and notably exhibit smaller expansion rates~\cite{Schlippert2020,Hensel2021}.   
While they readily attain expansion energies in the nanokelvin regime, delta-kick collimation techniques~\cite{Ammann1997PRL} can be employed for further reduction and values as low as \SI{38}{\pico\kelvin} have been demonstrated in a micro-gravity environment~\cite{Deppner21PRL}. 
However, the initial preparation of a BEC can significantly increase the experimental cycle time and neutralize a potential sensitivity gain.
Considerable efforts have been invested in overcoming this limitation by exploring rapid cooling schemes.
Source systems based on multi-layer atom-chips have been demonstrated as a convincing solution for magnetically trappable atoms and \Rb in particular~\cite{Rudolph15NJP,Becker2018}. 
Their strong confinement and high trap frequencies allows to realize BECs with more than \num{1e5} atoms and repetition rates on the order of \SI{1}{\hertz}, including less than \SI{500}{\milli\second} of evaporative cooling.
In a complementary approach, optical dipole traps (ODTs) are used when dealing with atoms insensitive to magnetic fields such as strontium or ytterbium or when experimental constraints do not allow for the implementation of chips. 
In these cases the issue is more severe as the inherent coupling of trap depth and trap frequencies intrinsically counteracts runaway evaporation at low intensity, resulting in evaporation sequences taking up to tens of seconds. 
Previous attempts to circumvent the corresponding scaling laws~\cite{OHara01PRA} include the use of movable lens systems~\cite{Kinoshita05PRAR}, time-averaged optical potentials~\cite{Roy2016PRA}, or hybrid approaches incorporating laser cooling on broad and narrow transitions~\cite{Stellmer13PRA} and direct laser cooling in the ODT~\cite{Urvoy19PRL}.
Most recently, machine learning techniques have also been implemented resulting in evaporation ramps with durations below \SI{200}{\milli\second}~\cite{Vendeiro2022PRR}.
Yet, most of these methods come with their own challenges regarding final atom number, expansion energy or condensate fraction.
Notably, optical setups which exhibit the same performance as chip traps are still lacking. 

In this work we present a novel scheme to enhance evaporative cooling of \K in an ODT.
Contrary to \Rb, \K offers the advantage of broad Feshbach resonances~\cite{Inouye98Nature} at low magnetic fields, which can be used to tailor interactions~\cite{DErrico07NJP}. 
By combining an initial trap compression with a dynamic tuning of the scattering length over two orders of magnitude and trapping frequencies along the evaporation ramp, we realize a nearly constant evaporation flux of \SI{3e5}{atoms\per\second} for an evaporation time ranging from \SI{170}{\milli\second} to \SI{2}{\second}.
Furthermore, tuning the scattering length to near zero thereafter offers a straightforward way to obtain a momentum spread close to the theoretical minimum, directly after release from the trapping potential.
Our source system improves the flux performance of an ODT setup to the one of state-of-the-art devices, enabling future sensor setups with superior sensitivity for a wider range of applications.
\section{Source system}
\subsection{Ensemble preparation}
\begin{figure}[t!]
    \centering
    \includegraphics[width = 0.49\textwidth]{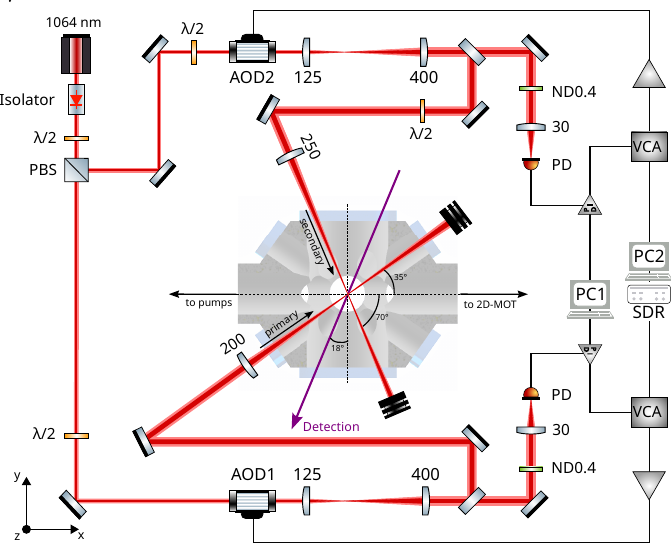}
    \caption{\textbf{Schematic representation of the crossed ODT setup.} 
    A single-frequency laser source [\textit{Coherent Mephisto MOPA}] is split into two independent beams, allowing for up to \SI{16}{\watt} per path.
    Beam waist were determined with a beamcam [DataRay TaperCamD-LCM], while the crossing angle of the beams within the chamber was measured with vertical imaging.
    Time-averaged potentials are implemented with AODs [\textit{AA Opto-Electronic DTSXY-400-1064}], allowing to independently modulate each beam's center position in the horizontal and vertical directions with a maximum modulation amplitude of \SI{1.5}{\milli\meter} [resp. \SI{1.8}{\milli\meter} for the secondary beam]. 
    In this work, only the horizontal axes of the AODs is modulated using a software defined radio [\textit{Ettus USRP X310}] to provide the waveform, while the vertical axes are driven with a constant frequency.
    For intensity stabilization, less than \SI{0.1}{\percent} of the optical power is detected by an amplified photodetector [\textit{Femto OE-200}] and used to control the diffraction efficiency of the AOD via a homebuild PID-controller together with a voltage controlled attenuate [\textit{MiniCircuits ZX73-2500-S+}].
    }
    \label{fig:ODT}
\end{figure}

Our ODT is created from two focused laser beams at a wavelength of \SI{1064}{\nano\meter}, crossing under an angle of \SI{74}{\degree} with natural beam waists of \SI{24.5\pm 1.6}{\micro\meter} [resp. \SI{30.3\pm 3.2}{\micro\meter}] for the primary [secondary] beam at the position of the atoms.
A simplified overview of our setup is presented in Fig.~\ref{fig:ODT}.

For loading we use an all-optical scheme without the need for a magnetic trap as intermediate step.
We first prepare \num{5e8} atoms at a temperature of \SI{5.93\pm 0.07}{\micro\kelvin} using gray molasses cooling on the \DOne line~\cite{Salomon13EPL} with the magneto-optical trapping setup and sequence as described in Ref.~\cite{Herbst2022PRA}. 
During the molasses step, we switch on the ODT and let the atoms fall freely through it afterwards, loading the trap within \SI{50}{\milli\second}.
Implementing time-averaged optical potentials by means of acousto-optic deflectors (AODs) we are not restricted to the natural beam waists, but rather realize a trap of harmonic shape and variable spatial size in the horizontal plane~\cite{Roy2016PRA}.
Due to the Ramsauer minimum of \K~\cite{Landini12PRA}, direct loading of traps deeper than \SI{400}{\micro\kelvin} is unfavorable and we always find the maximum loading efficiency at trap depths between \num{60} and \SI{80}{\micro\kelvin}. 
Instead of increasing the trap depth, we therefore use \SI{15.8}{\watt} of optical power per beam to increase the center-position modulation amplitude via the AODs, creating larger effective beam waists in the horizontal plane.
Extending the spatial overlap with the resulting pancake-shaped trap, we improve mode matching of ODT and molasses. 
The maximum atom number loaded into the ODT is achieved with equal horizontal modulation strokes of \SI{1.4}{\milli\meter} in the crossing region and a trap depth of \SI{65+-7}{\micro\kelvin}. 
In this configuration we load more than \num{2e7} atoms into the ODT.
Compared to the configuration without any spatial modulation, we hereby improve the loading performance by more than an order of magnitude.
Subsequently, we perform a multi-loop state preparation sequence as described previously in Ref.~\cite{Herbst2022PRA} and ultimately prepare a total of ~\num{14.8+-0.5 e6} atoms in $\ket{F=1,\mf=-1}$ with \num{10.7+-0.4 e6} atoms in the immediate crossing region at a temperature of \SI{7.14\pm 0.05}{\micro\kelvin}.
The high optical beam power together with the pancake shape also offers the advantage of achieving a high vertical trap frequency, resulting in a favorable initial phase space density (PSD) of \num{1.18\pm 0.59e-3} in the crossing region prior to evaporative cooling.
\subsection{Evaporative cooling}\label{sec:Evaporativ_cooling}
\begin{figure}[thb!]
    \begin{center}
    \includegraphics[width=1\columnwidth]{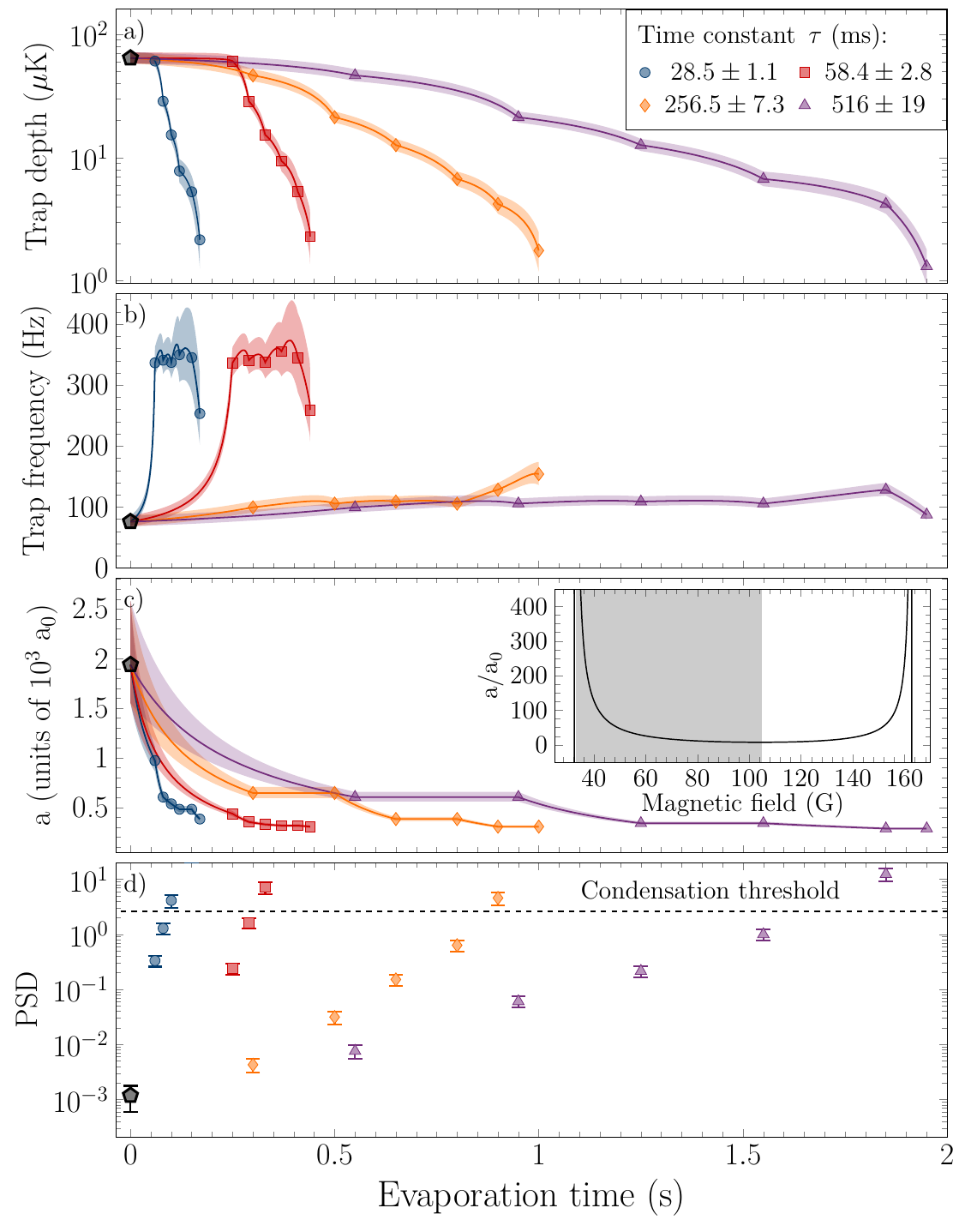}
    \caption{\textbf{Ramps optimized for evaporative cooling on different time scales with constant evaporative flux, together with the resulting phase space densities.}
    Along each of the four ramps, markers indicate start and end points of the six sections, which are individually optimized.
    Trap depths (a) and frequency curves (b) are obtained from a simulation of the confining potential. 
    Here, the error bands are estimated as the 2$\sigma$-uncertainty interval via error propagation, taking measurements of the beam waist, intensity, and geometry within the chamber into account. 
    For modelling the scattering length (c) we follow Ref.~\cite{Inouye98Nature}, together with the properties of the individual resonances as reported in Ref.~\cite{Tiemann2020PRR}. 
    Our experimentally determined magnetic field uncertainty is used to assign the error bands.
    The inset shows the overall behaviour of the scattering length in between the resonances at \SI{32.6}{\gauss} and \SI{162.8}{\gauss} for atoms in $\ket{F=1, m_\text{F}=-1}$. 
    Magnetic field values used in this work are highlighted by the shaded area.
    To determine the PSD (d), we perform atom number measurements at each marker position, averaging over 100 experimental cycles. 
    Temperatures are obtained from fitting the expansion velocity to TOF measurements of the ensemble size with data taken in-between \SI{1}{\milli\second} and \SI{30}{\milli\second} of free fall with \SI{1}{\milli\second} spacing and at least four measurements at each point in time.
    }
    \label{fig:Evaporation}
    \end{center}
\end{figure}
The optimization of our evaporative cooling sequence is based on the model provided in Ref.~\cite{Roy2016PRA}. 
The time needed for re-thermalization is inversely proportional to the elastic collision rate $\Gamma_{\text{el}}\propto N\bar{\omega}^3 a^2/T$ for atom number $N$, temperature $T$, geometric mean of the trapping frequencies $\bar{\omega}$ and scattering length $a$, which therefore limits the speed of evaporative cooling~\cite{Ketterle96AAMOP,Monroe1993PRL}. 
However, the efficiency of the process, and consequently the number of condensed particles realizable for given initial conditions, depends on the ratio $\beta$ of the evaporation rate to the remaining loss rates.
In the case of an ODT and evaporative cooling on time scales significantly faster than the trap lifetime, the three-body recombination rate $\Gamma_{\text{3b}}\propto N^2\bar{\omega}^6 a^4/T^{3}$ poses the dominant loss channel~\cite{Weber03PRL}.
Hence $\beta\approx\Gamma_\text{el}/\Gamma_\text{3b}\propto(N/\bar{\omega}^3)(T/a)^2$, allowing to optimize evaporation trajectories for either large atom numbers or high evaporation speed by choosing trapping frequencies and scattering length, accordingly. 

We optimize our evaporation sequence for a given total evaporation time towards the largest number of condensed particles, thus maximizing $\beta$, for a given value of $\Gamma_\text{el}$.
Extending beyond previous work~\cite{Herbst2022PRA}, we do so by not operating at a constant scattering length, but rather dynamically tuning the interactions within six linear ramps in coordination with the powers and spatial modulation of the optical beams.
Our optimized ramps in terms of trap depth, trap frequency, and scattering length together with the resulting phase space density are presented in Fig.~\ref{fig:Evaporation}.
For all ramps the initial configurations are identical and highlighted with a black pentagon. 
Each individual ramp duration is depicted in a different color.
Markers indicate the start and end points of the linear ramp sections, at which we also measure atom number and temperature to obtain the PSD. 
The optical power and modulation stroke are always chosen such that the trap depth decreases approximately exponentially with time constant $\tau$ as shown in Fig.~\ref{fig:Evaporation}a.
For the ramps with a total duration below \SI{1}{\second}, depicted in blue and red, we additionally perform a rapid compression as a first step prior to reducing the trap depth to increase the density of the ensemble.
This allows to achieve the high trap frequencies required to ensure the necessary evaporation rate for the short ramps early on, as shown in Fig.~\ref{fig:Evaporation}b. 
For the longer ramps, depicted in orange and purple, lower evaporation rates are sufficient and hence we choose frequencies comparable to the initial configuration. 
Especially for the short ramps an high initial scattering length of 
$1977^{+641}_{-387}\,$\unit{a_0} assists the compression, by maximizing the re-thermalization rate in the otherwise dilute sample.
By this we keep atom number loss associated with the related heating process from the compression $dT / dt = (\dot\omega / \omega)\, T$ to a minimum.  
In any case, we proceed by exponentially reducing the scattering length as shown in Fig.~\ref{fig:Evaporation}c. 
By doing so, we counteract the relative increase in losses from the temperature reduction by reducing the scattering length, since temperature and scattering length obey the same power law in $\beta$.
Here, evaporating in the vicinity of the broad Feshbach resonance at \SI{32.6}{\gauss} allows to precisely tune the interactions as needed. 
The behaviour of the scattering length for a wider range of magnetic fields is depicted in the inset of Fig.~\ref{fig:Evaporation}c, were the magnetic field range used in this work is highlighted as shaded grey area.  
For our shortest ramp we cross the phase transition after a total evaporation time of \SI{100}{\milli\second}, as indicated by the blue data point above the condensation threshold at $\zeta(3/2)\approx 2.612$ in Fig.\ref{fig:Evaporation}d, while requiring a total ramp length of \SI{170}{\milli\second} to achieve a quasi-pure condensate of \num{6.14 \pm 0.35 e4} particles.

\begin{figure}[tb!]
    \begin{center}
    \resizebox{0.99\columnwidth}{!}{\includegraphics[width=1\textwidth]{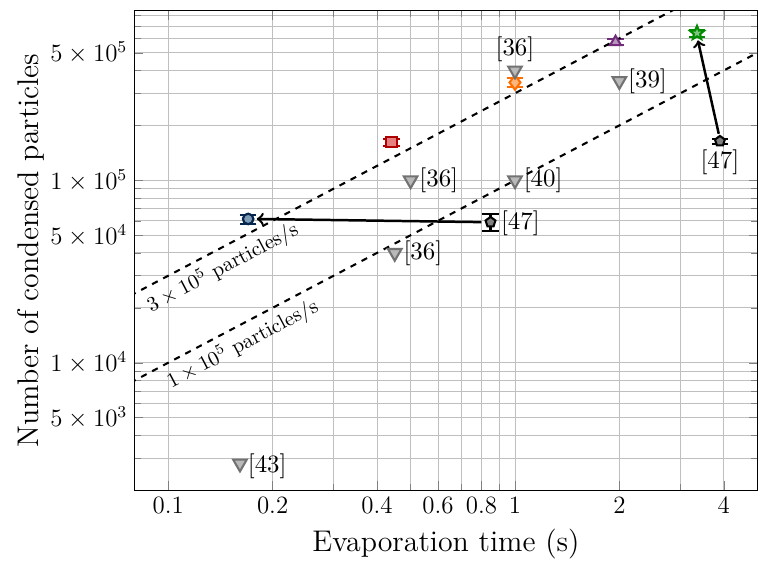}}
    \caption{\textbf{Final particle numbers in the condensate, against total evaporation time.} 
    Our previously achieved results with constant scattering length in a \SI{1960}{\nano\meter} trap are depicted with black pentagons, while our current results with variable scattering length in the \SI{1064}{\nano\meter} trap are highlighted, following the color and shape coding of Fig.~\ref{fig:Evaporation}.  
    The error bars are given by the standard deviation of 100 measurements of the particle number for each point.  
    For comparison, the performance of other fast BEC sources is depicted with grey upside-down triangles. 
    }
    \label{fig:Performance}
    \end{center}
\end{figure}

The evaporation performance for all ramps in terms of particle number is depicted in Fig.~\ref{fig:Performance} and compared to previously obtained results. 
Note that the color and shape coding resembles that one that was already used in Fig.~\ref{fig:Evaporation}.
The method presented here allows to achieve evaporation durations comparable to the machine-learning enhanced case~\cite{Vendeiro2022PRR}, but with a twenty-fold increase in atom number. 
Furthermore, unlike the high-flux chip source in Ref.~\cite{Rudolph15NJP}, we realize a nearly constant evaporation flux of \SI{3e5}{atoms\per\second} for evaporation times between \SI{170}{\milli\second} and \SI{2}{\second}.
For longer evaporation durations we find a significant reduction in flux as losses associated with the life time become non-negligible.
Nevertheless, we realize our largest BEC (green star) with \num{6.41 \pm 0.28 e5} atoms after an evaporation time of \SI{3.35}{\second} (ramp not shown in Fig.~\ref{fig:Evaporation}). 
Compared to our previous results with static scattering length in a \SI{1960}{\nano\meter}-trap, depicted as black pentagons~\cite{Herbst2022PRA}, we increased the speed of our fastest sequence by a factor of five while maintaining the particle number and improved the largest particle number by a factor of four, as marked by the black arrows.  
\begin{figure}[tbh]
    \begin{center}
    \resizebox{0.99\columnwidth}{!}{\includegraphics[width=1\textwidth]{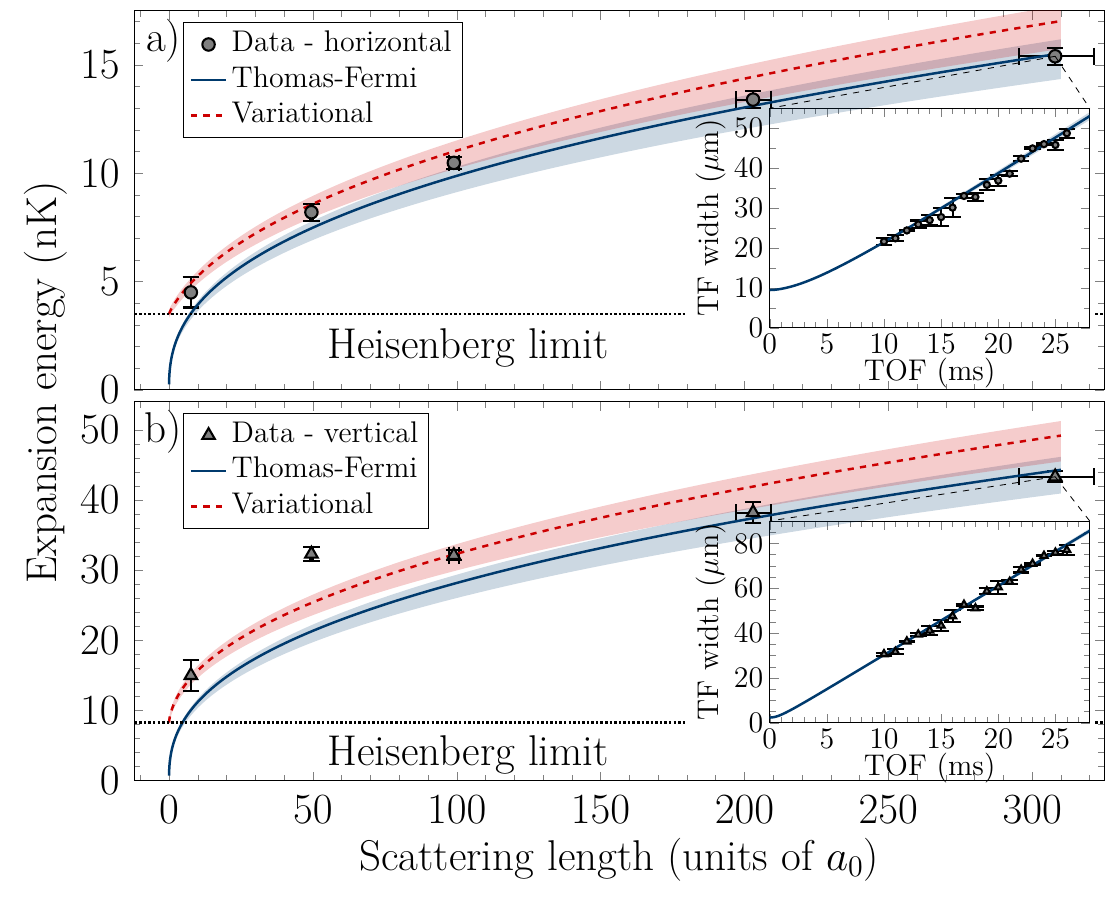}}
    \caption{\textbf{Expansion energy of the BEC at different scattering lengths in horizontal (a) and vertical (b) direction.}
    For each data point we perform a TOF series, determining the ensembles size in between \SI{10}{\milli\second} and \SI{26}{\milli\second} of free fall with \SI{1}{\milli\second} spacing and at least four measurements per point. 
    Contrary to the measurements performed in section~\ref{sec:Evaporativ_cooling} we do not measure below \SI{10}{\milli\second} to avoid the resolution limitation of our detection system. 
    The insets show the TOF-series for the data taken at \SI{308}{a_0}, which are used together with the measurements at \SI{203}{a_0} to determine the trap frequencies, by fitting a scaling approach, shown as solid lines, with the error band corresponding to the stated trap frequency error.
    The error bars of the energy measurements originate as one-sigma deviation from the fit error of the expansion velocity and from the magnetic field uncertainty for the scattering length.
    Error bands of the simulations are obtained via a Monte-Carlo method within the trap frequency interval.
    }
    \label{fig:LMS}
    \end{center}
\end{figure}
\subsection{Limited momentum spread}
We further exploit the tunability of the atomic interactions to minimize the expansion energy of the ensemble upon release from the trap.
Without interactions, the fundamental limit for the one dimensional momentum spread of a BEC released from a harmonic potential is determined by the oscillator length $a_\text{ho} = [\hbar/ m \omega ]^{1/2}$ with the corresponding energy $E = \hbar\omega/2 k_B$, when expressed in units of Kelvin. 
In experiments, larger energies are commonly observed as the non-vanishing mean field energy acts as a repulsive force upon release. 
Using a Feshbach resonance has already been demonstrated as a viable approach to reach the fundamental limit with BECs of cesium~\cite{Weber2003Science, Kraemer2004APB} and potassium~\cite{Roati2007PRL}.
We here confirm these results and measure the expansion energy for different scattering lengths by performing a least-square fit of the ensemble expansion based on TOF series (Fig.~\ref{fig:LMS}), which are useful for discussing the performance of an atom interferometer in the next section.

After creating the BEC using the \SI{1}{\second} long evaporation ramp, we adiabatically sweep the magnetic field and perform measurements at different scattering lengths between the $\ket{F=1, m_\text{F}=-1}$ resonances at \SI{32.6}{\gauss} and \SI{162.8}{\gauss}, as shown by grey shaded area in the inset of Fig.~\ref{fig:Evaporation}c. 
Notably, the broad minimum of \SI{7.59}{a_0} at \SI{104.1}{\gauss}, allows to approach an interaction free ensemble without the need for additional state transfers, but explicitly does not allow for zero or negative scattering lengths~\cite{DErrico07NJP}.
We then release the BEC from a trap with initial frequencies $\{\omega_x,\omega_y,\omega_z\}=2\pi\times\{145.5\pm 7.3, 11.5\pm 0.6, 342 \pm 17\}\,$Hz.
Here, trap frequencies are extracted from subsequent TOF measurements using a global fit on both datasets at $a=\SI{203\pm 6}{a_0}$ and $308^{+14}_{-13}\,$\unit{a_0}, allowing for a \SI{5}{\percent} error (Fig.~\ref{fig:LMS} insets).  
In this regime, we model the ensemble's dynamics during the TOF by solving a scaling approach, assuming a parabolic spatial distribution of the atomic density, consistent with the Thomas-Fermi (TF) approximation~\cite{Castin96PRL,Kagan97PRA}.
The RMS widths of the condensate density are given by $\rho_i(t)=\rho_i(0)\,\lambda_i(t)$ with scaling factor $\lambda_i(t)$ and $\rho_i(0)=R_i(0)/\sqrt{7}$, for $R_i(0)$ being the initial TF radius of the $i$-direction~\cite{Corgier_2020}. 
As the interactions decrease the TF approximation becomes less accurate, since the kinetic term becomes more relevant. 
Hence the initial width can no longer be described by the previous $\rho_i(0)$, which would shrink to zero and it becomes preferable to describe the BEC dynamics by a variational approach based on a Gaussian wave function with RMS width $\rho_i(t)$ for lower scattering length~\cite{Perez96PRL,Perez97PRA}. 

In Fig.~\ref{fig:LMS} we observe a good agreement between the TF-approximation (solid blue line) and the experimental data (black points and triangles), for scattering lengths above \SI{150}{a_0}, while the variational approach (dashed red lines) offers a better agreement below. 
At \SI{7.59+-0.01}{a_0} we find expansion energies of \SI{4.5+-0.7}{\nano\kelvin} [resp. \SI{15.0+-2.2}{\nano\kelvin}] in the horizontal [vertical] direction.
For a vanishing scattering length, e.g. realizable at \SI{43.7}{\gauss} for atoms in $\ket{F=1,\mf=0}$, the variational approach predicts a minimum expansion energy of \SI{3.5}{\nano\kelvin} [\SI{8.3}{\nano\kelvin}].
\section{Expected interferometer performance} 
We estimate the instability of an atom interferometer in Mach-Zehnder geometry for a setup utilizing our source configuration.
We follow the calculations performed in Ref.~\cite{Albers2022Commun} for a Raman beamsplitter with \SI{1.2}{\centi\meter} beam radius and a pulse duration of $t_\pi = \SI{15}{\micro\second}$.
At the standard quantum limit the instability of the interferometer after integration time $\tau$ is given by: 
\begin{equation*}
    \sigma(\tau) = \frac{1}{C\sqrt{N}k_\text{eff}T_\text{I}^2}\times\sqrt{\frac{t_\text{cycle}}{\tau}},
\end{equation*}
when neglecting the finite pulse duration.
Its scaling with $[t_\text{cycle}/N]^{1/2}$ results in the formerly stated requirement for a high atomic flux.
For our analysis we further divide the cycle time into the time in between interferometry pulses $T_\text{I}$, the evaporation time $t_\text{evap}$ and the remaining time $t_\text{prep}$ used for loading the ODT, state preparation and detection: $t_\text{cycle} = t_\text{prep}+t_\text{evap}+2T_\text{I}$.
Following Ref.~\cite{Loriani2019NJP}, we determine the contrast $C$ as the product of the excitation probabilities from the atom-light interactions. 
Here, the final expansion energy causes inhomogeneous Rabi frequencies, due to the velocity acceptance and intensity profile of the Raman beams. 
We calculate the resulting instability for different preparation times in combination with the evaporation times and the lowest expansion energy we experimentally demonstrated in the previous section.
Additionally, we compare the obtained results to the ones achievable with the chip trap.  

\begin{figure}[thb!]
    \begin{center}
    \includegraphics[width=1\columnwidth]{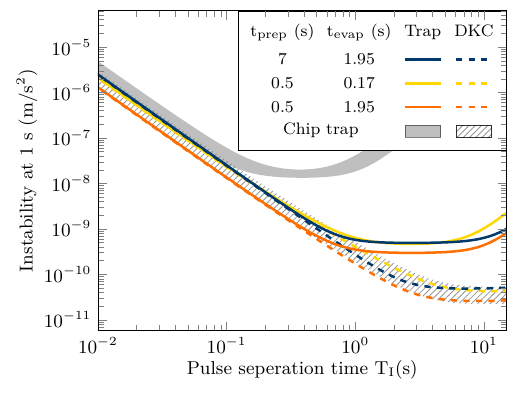}
    \caption{\textbf{Calculated instability of a Mach-Zehnder atom interferometer at the standard quantum limit using different source configurations with our evaporation sequence.}
    We show the expected instability for different pulse separation times for an immediate release from the trap (solid lines) and compare them to an interferometer after performing an additional matter-wave collimation to \SI{50}{\pico\kelvin} (dashed lines). 
    The shaded areas show the respective performance of a chip-based source system with values taken from Ref.~\cite{Rudolph15NJP} for the case of an immediate release from the trap and from Ref.~\cite{Deppner21PRL} for the DKC case.}
    \label{fig:Sensor}
    \end{center}
\end{figure}

Our experimental preparation time is limited to $t_\text{prep}=\SI{7}{\second}$ by MOT-loading and the time required for the final data transfer of the taken images. 
For our setup the total atomic flux $N/(t_\text{prep}+t_\text{evap})$ scales beneficially with the longer evaporation ramps since we keep $N/t_\text{evap}$ constant.
We hence choose $t_\text{evap} = \SI{1.95}{\second}$ for benchmarking. 
Here the calculation yields a minimum instability of \SI{5e-10}{\meter/\second\squared} at \SI{1}{\second} integration time, as indicated by the solid blue line in Fig.~\ref{fig:Sensor}.
When comparing to the accessible values by the chip trap directly after evaporative cooling, shown as the filled grey band, this corresponds to an improvement by over an order of magnitude, due to the smaller expansion energy of our setup which offsets our longer preparation time. 
Naturally, the results can be improved by a rapid MOT-loading scheme.
Preparation times below \SI{1}{\second} are readily achievable with an intense atomic source as provided by a 2D$^+$-MOT~\cite{Catani2006PRA,Chaudhuri06PRA}.
Even further reduction is possible when considering cryogenic sources~\cite{Lasner21PRA}.
A reasonable preparation time of $t_\text{prep}=\SI{500}{\milli\second}$ already yields an instability of \SI{3e-10}{\meter/\second\squared} as depicted by the solid orange line in Fig.~\ref{fig:Sensor}.
Additionally, such a setup allows to access the regime below \SI{1e-9}{\meter/\second\squared} even with our shortest evaporation ramp of $t_\text{evap} = \SI{170}{\milli\second}$ and enables sensors whose cycle time is entirely limited by the pulse separation time (yellow line).

Importantly, these results do not require additional matter-wave collimation when minimizing interactions. 
However, delta-kick collimation (DKC) techniques can be implemented to additionally reduce the expansion energy~\cite{Ammann1997PRL}. To estimate the expected performance in this regime, we consider the method that has already been demonstrated with an ODT on a long baseline, realizing expansion energies of \SI{50}{\pico\kelvin}~\cite{Kovachy15PRL}. 
For the chip based source system we consider the current record of \SI{38}{\pico\kelvin} which has been achieved in micro-gravity~\cite{Deppner21PRL}.
Since both setups now feature similar expansion energies and our evaporation sequences allow for similar cycle times we find the same instability regime below \SI{5e-11}{\meter/\second\squared} for both apparatuses, as shown by the overlap of the dashed lines with the area filled with a grey diagonal line pattern. 
\section{Outlook}
We have demonstrated a rapid, all-optical source system for large ensembles of quantum-degenerated \K, reduced its expansion energy by tuning atomic interactions and calculated the resulting instability of a Mach-Zehnder atom interferometer at the standard quantum limit.
Our analysis yields a superior performance for short cycle times and especially enables sensors which can prepare atomic ensembles well within the actual interferometry sequence. Hence the presented sequence allows to effectively reduce the dead time in-between measurements to zero, which has immediate applications in the field of hybrid inertial sensing, where a high data rate is required~\cite{Dutta16PRL}.

Moreover, for a delta-kick collimated ensemble we expect a performance similar to the best chip traps, which is of interest for experiments dedicated to high-accuracy measurements in fundamental physics.
Especially ground-based long baseline experiments benefit from the demonstrated method, e.g. the Very Long Baseline Atom Interferometer, which does not allow for the use of chip traps, due to the choice of atoms, scale of the device and optical access requirements~\cite{Hartwig15NJP,Schlippert20WS}.
In particular, we see applications for tests of quantum mechanics, e.g. in the context of the continuous spontaneous localization  model~\cite{Pearle1989PRA,Ghirardi1990PRA, Kovachy15Nature}.
Here, the sensitivity scales with the third power of the number of condensed particles and a rapid ensemble preparation together with a minimal final scattering length is required~\cite{Schrinski23PRA}.
Moreover, realizing interaction-free BECs in free fall opens up the possibility of another class of non-interferometric tests, probing deviations from the uncertainty principle due to additional heating processes~\cite{Bassi2013RMP,Carlesso.2022}.

Beyond \K, the evaporation methods demonstrated can also be applied to source systems of other atomic species.
For rubidium, suitable magnetic Feshbach resonances exist~\cite{Roberts1998PRL,Marte2002PRL}, but their narrower width and higher magnetic field strength make them technologically more challenging for improving evaporative cooling. 
For strontium and ytterbium, magnetic resonances are not available due to the non-degenerate nature of their ground state.
In this case, optical Feshbach resonances (OFR), which modulate interatomic interactions by coupling two colliding atoms to a bound molecular state, have recently gained interest, due to their potential applications in molecule formation~\cite{Koch2005PRL,Koch2008PRA,PhysRevLett.111.150402,PhysRevLett.116.043202}.
While broad OFRs are accompanied by high losses caused by the spontaneous decay of the excited molecular state~\cite{Theis2004PRL}, narrow linewidth resonances of the forbidden $^{1}$S$_{0}$ - $^{3}$P$_{1}$ intercombination transition suppress this behavior and have been used to efficiently change the atomic scattering length across large intervals~\cite{Ciuryo2005PRL,Enomoto2008PRL,Blatt11PRL}.
Finally, a resonance suitable for thermalization has been identified for $^{88}$Sr~\cite{Zelevinsky2006PRL,Blatt11PRL}, offering prospects for direct evaporative cooling and its optimization using our method.
\begin{acknowledgments}
We thank Jan Rudolph for fruitful discussions and helpful comments regarding the general scope of the manuscript, the performance of evaporative cooling and for thorough proof reading.
This work is funded by the Federal Ministry of Education and Research (BMBF) through the funding program Photonics Research Germany under contract number 13N14875 
and supported by the ``ADI 2022'' project founded by the IDEX Paris-Saclay, ANR-11-IDEX-0003-02.
The authors further acknowledge support by the German Space Agency (DLR) with funds
provided by the Federal Ministry for Economic Affairs and Climate Action due to an enactment of the German Bundestag under Grant No. DLR 50WM2041 (PRIMUS-IV), 50WM2253A (AI-Quadrat), and by the Deutsche Forschungsgemeinschaft (DFG, German Research Foundation)–Project-ID 274200144–the SFB 1227 DQ-mat within Project No.~A05 and ~B07 and under Germany’s Excellence Strategy—EXC-2123 QuantumFrontiers—Project-ID 390837967.
\end{acknowledgments}
\bibliography{main}

\begin{thebibliography}{82}%
\makeatletter
\providecommand \@ifxundefined [1]{%
 \@ifx{#1\undefined}
}%
\providecommand \@ifnum [1]{%
 \ifnum #1\expandafter \@firstoftwo
 \else \expandafter \@secondoftwo
 \fi
}%
\providecommand \@ifx [1]{%
 \ifx #1\expandafter \@firstoftwo
 \else \expandafter \@secondoftwo
 \fi
}%
\providecommand \natexlab [1]{#1}%
\providecommand \enquote  [1]{``#1''}%
\providecommand \bibnamefont  [1]{#1}%
\providecommand \bibfnamefont [1]{#1}%
\providecommand \citenamefont [1]{#1}%
\providecommand \href@noop [0]{\@secondoftwo}%
\providecommand \href [0]{\begingroup \@sanitize@url \@href}%
\providecommand \@href[1]{\@@startlink{#1}\@@href}%
\providecommand \@@href[1]{\endgroup#1\@@endlink}%
\providecommand \@sanitize@url [0]{\catcode `\\12\catcode `\$12\catcode
  `\&12\catcode `\#12\catcode `\^12\catcode `\_12\catcode `\%12\relax}%
\providecommand \@@startlink[1]{}%
\providecommand \@@endlink[0]{}%
\providecommand \url  [0]{\begingroup\@sanitize@url \@url }%
\providecommand \@url [1]{\endgroup\@href {#1}{\urlprefix }}%
\providecommand \urlprefix  [0]{URL }%
\providecommand \Eprint [0]{\href }%
\providecommand \doibase [0]{http://dx.doi.org/}%
\providecommand \selectlanguage [0]{\@gobble}%
\providecommand \bibinfo  [0]{\@secondoftwo}%
\providecommand \bibfield  [0]{\@secondoftwo}%
\providecommand \translation [1]{[#1]}%
\providecommand \BibitemOpen [0]{}%
\providecommand \bibitemStop [0]{}%
\providecommand \bibitemNoStop [0]{.\EOS\space}%
\providecommand \EOS [0]{\spacefactor3000\relax}%
\providecommand \BibitemShut  [1]{\csname bibitem#1\endcsname}%
\let\auto@bib@innerbib\@empty
\bibitem [{\citenamefont {Kasevich}\ and\ \citenamefont
  {Chu}(1991)}]{Kasevich91PRL}%
  \BibitemOpen
  \bibfield  {author} {\bibinfo {author} {\bibfnamefont {M.}~\bibnamefont
  {Kasevich}}\ and\ \bibinfo {author} {\bibfnamefont {S.}~\bibnamefont {Chu}},\
  }\href {\doibase 10.1103/PhysRevLett.67.181} {\bibfield  {journal} {\bibinfo
  {journal} {Phys. Rev. Lett.}\ }\textbf {\bibinfo {volume} {67}},\ \bibinfo
  {pages} {181} (\bibinfo {year} {1991})}\BibitemShut {NoStop}%
\bibitem [{\citenamefont {Riehle}\ \emph {et~al.}(1991)\citenamefont {Riehle},
  \citenamefont {Kisters}, \citenamefont {Witte}, \citenamefont {Helmcke},\
  and\ \citenamefont {Bord\'e}}]{Riehle91PRL}%
  \BibitemOpen
  \bibfield  {author} {\bibinfo {author} {\bibfnamefont {F.}~\bibnamefont
  {Riehle}}, \bibinfo {author} {\bibfnamefont {T.}~\bibnamefont {Kisters}},
  \bibinfo {author} {\bibfnamefont {A.}~\bibnamefont {Witte}}, \bibinfo
  {author} {\bibfnamefont {J.}~\bibnamefont {Helmcke}}, \ and\ \bibinfo
  {author} {\bibfnamefont {C.~J.}\ \bibnamefont {Bord\'e}},\ }\href {\doibase
  10.1103/PhysRevLett.67.177} {\bibfield  {journal} {\bibinfo  {journal} {Phys.
  Rev. Lett.}\ }\textbf {\bibinfo {volume} {67}},\ \bibinfo {pages} {177}
  (\bibinfo {year} {1991})}\BibitemShut {NoStop}%
\bibitem [{\citenamefont {Kasevich}\ and\ \citenamefont
  {Chu}(1992)}]{Kasevich92APB}%
  \BibitemOpen
  \bibfield  {author} {\bibinfo {author} {\bibfnamefont {M.}~\bibnamefont
  {Kasevich}}\ and\ \bibinfo {author} {\bibfnamefont {S.}~\bibnamefont {Chu}},\
  }\href {\doibase 10.1007/BF00325375} {\bibfield  {journal} {\bibinfo
  {journal} {Applied Physics B Photophysics and Laser Chemistry}\ }\textbf
  {\bibinfo {volume} {54}},\ \bibinfo {pages} {321} (\bibinfo {year}
  {1992})}\BibitemShut {NoStop}%
\bibitem [{\citenamefont {Cronin}\ \emph {et~al.}(2009)\citenamefont {Cronin},
  \citenamefont {Schmiedmayer},\ and\ \citenamefont {Pritchard}}]{Cronin09RMP}%
  \BibitemOpen
  \bibfield  {author} {\bibinfo {author} {\bibfnamefont {A.~D.}\ \bibnamefont
  {Cronin}}, \bibinfo {author} {\bibfnamefont {J.}~\bibnamefont
  {Schmiedmayer}}, \ and\ \bibinfo {author} {\bibfnamefont {D.~E.}\
  \bibnamefont {Pritchard}},\ }\href {\doibase 10.1103/RevModPhys.81.1051}
  {\bibfield  {journal} {\bibinfo  {journal} {Rev. Mod. Phys.}\ }\textbf
  {\bibinfo {volume} {81}},\ \bibinfo {pages} {1051} (\bibinfo {year}
  {2009})}\BibitemShut {NoStop}%
\bibitem [{\citenamefont {Gustavson}\ \emph {et~al.}(1997)\citenamefont
  {Gustavson}, \citenamefont {Bouyer},\ and\ \citenamefont
  {Kasevich}}]{Gustavson97PRL}%
  \BibitemOpen
  \bibfield  {author} {\bibinfo {author} {\bibfnamefont {T.~L.}\ \bibnamefont
  {Gustavson}}, \bibinfo {author} {\bibfnamefont {P.}~\bibnamefont {Bouyer}}, \
  and\ \bibinfo {author} {\bibfnamefont {M.~A.}\ \bibnamefont {Kasevich}},\
  }\href {\doibase 10.1103/PhysRevLett.78.2046} {\bibfield  {journal} {\bibinfo
   {journal} {Phys. Rev. Lett.}\ }\textbf {\bibinfo {volume} {78}},\ \bibinfo
  {pages} {2046} (\bibinfo {year} {1997})}\BibitemShut {NoStop}%
\bibitem [{\citenamefont {Canuel}\ \emph {et~al.}(2006)\citenamefont {Canuel},
  \citenamefont {Leduc}, \citenamefont {Holleville}, \citenamefont {Gauguet},
  \citenamefont {Fils}, \citenamefont {Virdis}, \citenamefont {Clairon},
  \citenamefont {Dimarcq}, \citenamefont {Bord\'e}, \citenamefont {Landragin},\
  and\ \citenamefont {Bouyer}}]{Canuel06PRL}%
  \BibitemOpen
  \bibfield  {author} {\bibinfo {author} {\bibfnamefont {B.}~\bibnamefont
  {Canuel}}, \bibinfo {author} {\bibfnamefont {F.}~\bibnamefont {Leduc}},
  \bibinfo {author} {\bibfnamefont {D.}~\bibnamefont {Holleville}}, \bibinfo
  {author} {\bibfnamefont {A.}~\bibnamefont {Gauguet}}, \bibinfo {author}
  {\bibfnamefont {J.}~\bibnamefont {Fils}}, \bibinfo {author} {\bibfnamefont
  {A.}~\bibnamefont {Virdis}}, \bibinfo {author} {\bibfnamefont
  {A.}~\bibnamefont {Clairon}}, \bibinfo {author} {\bibfnamefont
  {N.}~\bibnamefont {Dimarcq}}, \bibinfo {author} {\bibfnamefont {C.~J.}\
  \bibnamefont {Bord\'e}}, \bibinfo {author} {\bibfnamefont {A.}~\bibnamefont
  {Landragin}}, \ and\ \bibinfo {author} {\bibfnamefont {P.}~\bibnamefont
  {Bouyer}},\ }\href {\doibase 10.1103/PhysRevLett.97.010402} {\bibfield
  {journal} {\bibinfo  {journal} {Phys. Rev. Lett.}\ }\textbf {\bibinfo
  {volume} {97}},\ \bibinfo {pages} {010402} (\bibinfo {year}
  {2006})}\BibitemShut {NoStop}%
\bibitem [{\citenamefont {Dickerson}\ \emph {et~al.}(2013)\citenamefont
  {Dickerson}, \citenamefont {Hogan}, \citenamefont {Sugarbaker}, \citenamefont
  {Johnson},\ and\ \citenamefont {Kasevich}}]{Dickerson13PRL}%
  \BibitemOpen
  \bibfield  {author} {\bibinfo {author} {\bibfnamefont {S.~M.}\ \bibnamefont
  {Dickerson}}, \bibinfo {author} {\bibfnamefont {J.~M.}\ \bibnamefont
  {Hogan}}, \bibinfo {author} {\bibfnamefont {A.}~\bibnamefont {Sugarbaker}},
  \bibinfo {author} {\bibfnamefont {D.~M.~S.}\ \bibnamefont {Johnson}}, \ and\
  \bibinfo {author} {\bibfnamefont {M.~A.}\ \bibnamefont {Kasevich}},\ }\href
  {\doibase 10.1103/PhysRevLett.111.083001} {\bibfield  {journal} {\bibinfo
  {journal} {Phys. Rev. Lett.}\ }\textbf {\bibinfo {volume} {111}},\ \bibinfo
  {pages} {083001} (\bibinfo {year} {2013})}\BibitemShut {NoStop}%
\bibitem [{\citenamefont {Dutta}\ \emph {et~al.}(2016)\citenamefont {Dutta},
  \citenamefont {Savoie}, \citenamefont {Fang}, \citenamefont {Venon},
  \citenamefont {Garrido~Alzar}, \citenamefont {Geiger},\ and\ \citenamefont
  {Landragin}}]{Dutta16PRL}%
  \BibitemOpen
  \bibfield  {author} {\bibinfo {author} {\bibfnamefont {I.}~\bibnamefont
  {Dutta}}, \bibinfo {author} {\bibfnamefont {D.}~\bibnamefont {Savoie}},
  \bibinfo {author} {\bibfnamefont {B.}~\bibnamefont {Fang}}, \bibinfo {author}
  {\bibfnamefont {B.}~\bibnamefont {Venon}}, \bibinfo {author} {\bibfnamefont
  {C.~L.}\ \bibnamefont {Garrido~Alzar}}, \bibinfo {author} {\bibfnamefont
  {R.}~\bibnamefont {Geiger}}, \ and\ \bibinfo {author} {\bibfnamefont
  {A.}~\bibnamefont {Landragin}},\ }\href {\doibase
  10.1103/PhysRevLett.116.183003} {\bibfield  {journal} {\bibinfo  {journal}
  {Phys. Rev. Lett.}\ }\textbf {\bibinfo {volume} {116}},\ \bibinfo {pages}
  {183003} (\bibinfo {year} {2016})}\BibitemShut {NoStop}%
\bibitem [{\citenamefont {Savoie}\ \emph {et~al.}(2018)\citenamefont {Savoie},
  \citenamefont {Altorio}, \citenamefont {Fang}, \citenamefont {Sidorenkov},
  \citenamefont {Geiger},\ and\ \citenamefont {Landragin}}]{Savoie2018SciAdv}%
  \BibitemOpen
  \bibfield  {author} {\bibinfo {author} {\bibfnamefont {D.}~\bibnamefont
  {Savoie}}, \bibinfo {author} {\bibfnamefont {M.}~\bibnamefont {Altorio}},
  \bibinfo {author} {\bibfnamefont {B.}~\bibnamefont {Fang}}, \bibinfo {author}
  {\bibfnamefont {L.~A.}\ \bibnamefont {Sidorenkov}}, \bibinfo {author}
  {\bibfnamefont {R.}~\bibnamefont {Geiger}}, \ and\ \bibinfo {author}
  {\bibfnamefont {A.}~\bibnamefont {Landragin}},\ }\href {\doibase
  10.1126/sciadv.aau7948} {\bibfield  {journal} {\bibinfo  {journal} {Science
  Advances}\ }\textbf {\bibinfo {volume} {4}},\ \bibinfo {pages} {eaau7948}
  (\bibinfo {year} {2018})}\BibitemShut {NoStop}%
\bibitem [{\citenamefont {Schlippert}\ \emph {et~al.}(2014)\citenamefont
  {Schlippert}, \citenamefont {Hartwig}, \citenamefont {Albers}, \citenamefont
  {Richardson}, \citenamefont {Schubert}, \citenamefont {Roura}, \citenamefont
  {Schleich}, \citenamefont {Ertmer},\ and\ \citenamefont
  {Rasel}}]{Schlippert14PRL}%
  \BibitemOpen
  \bibfield  {author} {\bibinfo {author} {\bibfnamefont {D.}~\bibnamefont
  {Schlippert}}, \bibinfo {author} {\bibfnamefont {J.}~\bibnamefont {Hartwig}},
  \bibinfo {author} {\bibfnamefont {H.}~\bibnamefont {Albers}}, \bibinfo
  {author} {\bibfnamefont {L.~L.}\ \bibnamefont {Richardson}}, \bibinfo
  {author} {\bibfnamefont {C.}~\bibnamefont {Schubert}}, \bibinfo {author}
  {\bibfnamefont {A.}~\bibnamefont {Roura}}, \bibinfo {author} {\bibfnamefont
  {W.~P.}\ \bibnamefont {Schleich}}, \bibinfo {author} {\bibfnamefont
  {W.}~\bibnamefont {Ertmer}}, \ and\ \bibinfo {author} {\bibfnamefont {E.~M.}\
  \bibnamefont {Rasel}},\ }\href {\doibase 10.1103/PhysRevLett.112.203002}
  {\bibfield  {journal} {\bibinfo  {journal} {Phys. Rev. Lett.}\ }\textbf
  {\bibinfo {volume} {112}},\ \bibinfo {pages} {203002} (\bibinfo {year}
  {2014})}\BibitemShut {NoStop}%
\bibitem [{\citenamefont {Tarallo}\ \emph {et~al.}(2014)\citenamefont
  {Tarallo}, \citenamefont {Mazzoni}, \citenamefont {Poli}, \citenamefont
  {Sutyrin}, \citenamefont {Zhang},\ and\ \citenamefont
  {Tino}}]{Tarallo2014PRL}%
  \BibitemOpen
  \bibfield  {author} {\bibinfo {author} {\bibfnamefont {M.~G.}\ \bibnamefont
  {Tarallo}}, \bibinfo {author} {\bibfnamefont {T.}~\bibnamefont {Mazzoni}},
  \bibinfo {author} {\bibfnamefont {N.}~\bibnamefont {Poli}}, \bibinfo {author}
  {\bibfnamefont {D.~V.}\ \bibnamefont {Sutyrin}}, \bibinfo {author}
  {\bibfnamefont {X.}~\bibnamefont {Zhang}}, \ and\ \bibinfo {author}
  {\bibfnamefont {G.~M.}\ \bibnamefont {Tino}},\ }\href {\doibase
  10.1103/PhysRevLett.113.023005} {\bibfield  {journal} {\bibinfo  {journal}
  {Phys. Rev. Lett.}\ }\textbf {\bibinfo {volume} {113}},\ \bibinfo {pages}
  {023005} (\bibinfo {year} {2014})}\BibitemShut {NoStop}%
\bibitem [{\citenamefont {Albers}\ \emph {et~al.}(2020)\citenamefont {Albers},
  \citenamefont {Herbst}, \citenamefont {Richardson}, \citenamefont {Heine},
  \citenamefont {Nath}, \citenamefont {Hartwig}, \citenamefont {Schubert},
  \citenamefont {Vogt}, \citenamefont {Woltmann}, \citenamefont {Lämmerzahl},
  \citenamefont {Herrmann}, \citenamefont {Ertmer}, \citenamefont {Rasel},\
  and\ \citenamefont {Schlippert}}]{Albers2020EPJD}%
  \BibitemOpen
  \bibfield  {author} {\bibinfo {author} {\bibfnamefont {H.}~\bibnamefont
  {Albers}}, \bibinfo {author} {\bibfnamefont {A.}~\bibnamefont {Herbst}},
  \bibinfo {author} {\bibfnamefont {L.~L.}\ \bibnamefont {Richardson}},
  \bibinfo {author} {\bibfnamefont {H.}~\bibnamefont {Heine}}, \bibinfo
  {author} {\bibfnamefont {D.}~\bibnamefont {Nath}}, \bibinfo {author}
  {\bibfnamefont {J.}~\bibnamefont {Hartwig}}, \bibinfo {author} {\bibfnamefont
  {C.}~\bibnamefont {Schubert}}, \bibinfo {author} {\bibfnamefont
  {C.}~\bibnamefont {Vogt}}, \bibinfo {author} {\bibfnamefont {M.}~\bibnamefont
  {Woltmann}}, \bibinfo {author} {\bibfnamefont {C.}~\bibnamefont
  {Lämmerzahl}}, \bibinfo {author} {\bibfnamefont {S.}~\bibnamefont
  {Herrmann}}, \bibinfo {author} {\bibfnamefont {W.}~\bibnamefont {Ertmer}},
  \bibinfo {author} {\bibfnamefont {E.~M.}\ \bibnamefont {Rasel}}, \ and\
  \bibinfo {author} {\bibfnamefont {D.}~\bibnamefont {Schlippert}},\ }\href
  {\doibase 10.1140/epjd/e2020-10132-6} {\bibfield  {journal} {\bibinfo
  {journal} {The European Physical Journal D}\ }\textbf {\bibinfo {volume}
  {74}} (\bibinfo {year} {2020}),\ 10.1140/epjd/e2020-10132-6}\BibitemShut
  {NoStop}%
\bibitem [{\citenamefont {Asenbaum}\ \emph {et~al.}(2020)\citenamefont
  {Asenbaum}, \citenamefont {Overstreet}, \citenamefont {Kim}, \citenamefont
  {Curti},\ and\ \citenamefont {Kasevich}}]{Asenbaum2020PRL}%
  \BibitemOpen
  \bibfield  {author} {\bibinfo {author} {\bibfnamefont {P.}~\bibnamefont
  {Asenbaum}}, \bibinfo {author} {\bibfnamefont {C.}~\bibnamefont
  {Overstreet}}, \bibinfo {author} {\bibfnamefont {M.}~\bibnamefont {Kim}},
  \bibinfo {author} {\bibfnamefont {J.}~\bibnamefont {Curti}}, \ and\ \bibinfo
  {author} {\bibfnamefont {M.~A.}\ \bibnamefont {Kasevich}},\ }\href {\doibase
  10.1103/PhysRevLett.125.191101} {\bibfield  {journal} {\bibinfo  {journal}
  {Phys. Rev. Lett.}\ }\textbf {\bibinfo {volume} {125}},\ \bibinfo {pages}
  {191101} (\bibinfo {year} {2020})}\BibitemShut {NoStop}%
\bibitem [{\citenamefont {Bassi}\ \emph {et~al.}(2013)\citenamefont {Bassi},
  \citenamefont {Lochan}, \citenamefont {Satin}, \citenamefont {Singh},\ and\
  \citenamefont {Ulbricht}}]{Bassi2013RMP}%
  \BibitemOpen
  \bibfield  {author} {\bibinfo {author} {\bibfnamefont {A.}~\bibnamefont
  {Bassi}}, \bibinfo {author} {\bibfnamefont {K.}~\bibnamefont {Lochan}},
  \bibinfo {author} {\bibfnamefont {S.}~\bibnamefont {Satin}}, \bibinfo
  {author} {\bibfnamefont {T.~P.}\ \bibnamefont {Singh}}, \ and\ \bibinfo
  {author} {\bibfnamefont {H.}~\bibnamefont {Ulbricht}},\ }\href {\doibase
  10.1103/RevModPhys.85.471} {\bibfield  {journal} {\bibinfo  {journal} {Rev.
  Mod. Phys.}\ }\textbf {\bibinfo {volume} {85}},\ \bibinfo {pages} {471}
  (\bibinfo {year} {2013})}\BibitemShut {NoStop}%
\bibitem [{\citenamefont {Kovachy}\ \emph
  {et~al.}(2015{\natexlab{a}})\citenamefont {Kovachy}, \citenamefont
  {Asenbaum}, \citenamefont {Overstreet}, \citenamefont {Donnelly},
  \citenamefont {Dickerson}, \citenamefont {Sugarbaker}, \citenamefont
  {Hogan},\ and\ \citenamefont {Kasevich}}]{Kovachy15Nature}%
  \BibitemOpen
  \bibfield  {author} {\bibinfo {author} {\bibfnamefont {T.}~\bibnamefont
  {Kovachy}}, \bibinfo {author} {\bibfnamefont {P.}~\bibnamefont {Asenbaum}},
  \bibinfo {author} {\bibfnamefont {C.}~\bibnamefont {Overstreet}}, \bibinfo
  {author} {\bibfnamefont {C.~A.}\ \bibnamefont {Donnelly}}, \bibinfo {author}
  {\bibfnamefont {S.~M.}\ \bibnamefont {Dickerson}}, \bibinfo {author}
  {\bibfnamefont {A.}~\bibnamefont {Sugarbaker}}, \bibinfo {author}
  {\bibfnamefont {J.~M.}\ \bibnamefont {Hogan}}, \ and\ \bibinfo {author}
  {\bibfnamefont {M.~A.}\ \bibnamefont {Kasevich}},\ }\href {\doibase
  10.1038/nature16155} {\bibfield  {journal} {\bibinfo  {journal} {Nature}\
  }\textbf {\bibinfo {volume} {528}},\ \bibinfo {pages} {530} (\bibinfo {year}
  {2015}{\natexlab{a}})}\BibitemShut {NoStop}%
\bibitem [{\citenamefont {Carlesso}\ \emph
  {et~al.}(2022{\natexlab{a}})\citenamefont {Carlesso}, \citenamefont {Donadi},
  \citenamefont {Ferialdi}, \citenamefont {Paternostro}, \citenamefont
  {Ulbricht},\ and\ \citenamefont {Bassi}}]{Carlesso2022NatPhys}%
  \BibitemOpen
  \bibfield  {author} {\bibinfo {author} {\bibfnamefont {M.}~\bibnamefont
  {Carlesso}}, \bibinfo {author} {\bibfnamefont {S.}~\bibnamefont {Donadi}},
  \bibinfo {author} {\bibfnamefont {L.}~\bibnamefont {Ferialdi}}, \bibinfo
  {author} {\bibfnamefont {M.}~\bibnamefont {Paternostro}}, \bibinfo {author}
  {\bibfnamefont {H.}~\bibnamefont {Ulbricht}}, \ and\ \bibinfo {author}
  {\bibfnamefont {A.}~\bibnamefont {Bassi}},\ }\href {\doibase
  10.1038/s41567-021-01489-5} {\bibfield  {journal} {\bibinfo  {journal}
  {Nature Physics}\ }\textbf {\bibinfo {volume} {18}},\ \bibinfo {pages} {243}
  (\bibinfo {year} {2022}{\natexlab{a}})}\BibitemShut {NoStop}%
\bibitem [{\citenamefont {Schrinski}\ \emph {et~al.}(2023)\citenamefont
  {Schrinski}, \citenamefont {Haslinger}, \citenamefont {Schmiedmayer},
  \citenamefont {Hornberger},\ and\ \citenamefont
  {Nimmrichter}}]{Schrinski23PRA}%
  \BibitemOpen
  \bibfield  {author} {\bibinfo {author} {\bibfnamefont {B.}~\bibnamefont
  {Schrinski}}, \bibinfo {author} {\bibfnamefont {P.}~\bibnamefont
  {Haslinger}}, \bibinfo {author} {\bibfnamefont {J.}~\bibnamefont
  {Schmiedmayer}}, \bibinfo {author} {\bibfnamefont {K.}~\bibnamefont
  {Hornberger}}, \ and\ \bibinfo {author} {\bibfnamefont {S.}~\bibnamefont
  {Nimmrichter}},\ }\href {\doibase 10.1103/PhysRevA.107.043320} {\bibfield
  {journal} {\bibinfo  {journal} {Phys. Rev. A}\ }\textbf {\bibinfo {volume}
  {107}},\ \bibinfo {pages} {043320} (\bibinfo {year} {2023})}\BibitemShut
  {NoStop}%
\bibitem [{\citenamefont {Rosi}\ \emph {et~al.}(2014)\citenamefont {Rosi},
  \citenamefont {Sorrentino}, \citenamefont {Cacciapuoti}, \citenamefont
  {Prevedelli},\ and\ \citenamefont {Tino}}]{Rosi14Nature}%
  \BibitemOpen
  \bibfield  {author} {\bibinfo {author} {\bibfnamefont {G.}~\bibnamefont
  {Rosi}}, \bibinfo {author} {\bibfnamefont {F.}~\bibnamefont {Sorrentino}},
  \bibinfo {author} {\bibfnamefont {L.}~\bibnamefont {Cacciapuoti}}, \bibinfo
  {author} {\bibfnamefont {M.}~\bibnamefont {Prevedelli}}, \ and\ \bibinfo
  {author} {\bibfnamefont {G.~M.}\ \bibnamefont {Tino}},\ }\href {\doibase
  10.1038/nature13433} {\bibfield  {journal} {\bibinfo  {journal} {Nature}\
  }\textbf {\bibinfo {volume} {510}},\ \bibinfo {pages} {518} (\bibinfo {year}
  {2014})}\BibitemShut {NoStop}%
\bibitem [{\citenamefont {Parker}\ \emph {et~al.}()\citenamefont {Parker},
  \citenamefont {Yu}, \citenamefont {Zhong}, \citenamefont {Estey},\ and\
  \citenamefont {M\"uller}}]{Parker2018Science}%
  \BibitemOpen
  \bibfield  {author} {\bibinfo {author} {\bibfnamefont {R.~H.}\ \bibnamefont
  {Parker}}, \bibinfo {author} {\bibfnamefont {C.}~\bibnamefont {Yu}}, \bibinfo
  {author} {\bibfnamefont {W.}~\bibnamefont {Zhong}}, \bibinfo {author}
  {\bibfnamefont {B.}~\bibnamefont {Estey}}, \ and\ \bibinfo {author}
  {\bibfnamefont {H.}~\bibnamefont {M\"uller}},\ }\href {\doibase
  10.1126/science.aap7706} {\ \textbf {\bibinfo {volume} {360}},\ \bibinfo
  {pages} {191}}\BibitemShut {NoStop}%
\bibitem [{\citenamefont {Morel}\ \emph {et~al.}(2020)\citenamefont {Morel},
  \citenamefont {Yao}, \citenamefont {Clad{\'e}},\ and\ \citenamefont
  {Guellati-Kh{\'e}lifa}}]{Morel2020Nature}%
  \BibitemOpen
  \bibfield  {author} {\bibinfo {author} {\bibfnamefont {L.}~\bibnamefont
  {Morel}}, \bibinfo {author} {\bibfnamefont {Z.}~\bibnamefont {Yao}}, \bibinfo
  {author} {\bibfnamefont {P.}~\bibnamefont {Clad{\'e}}}, \ and\ \bibinfo
  {author} {\bibfnamefont {S.}~\bibnamefont {Guellati-Kh{\'e}lifa}},\ }\href
  {\doibase 10.1038/s41586-020-2964-7} {\bibfield  {journal} {\bibinfo
  {journal} {Nature}\ }\textbf {\bibinfo {volume} {588}},\ \bibinfo {pages}
  {61} (\bibinfo {year} {2020})}\BibitemShut {NoStop}%
\bibitem [{\citenamefont {El-Neaj}\ \emph {et~al.}(2020)\citenamefont
  {El-Neaj}, \citenamefont {Alpigiani}, \citenamefont {Amairi-Pyka},
  \citenamefont {Ara{\'u}jo}, \citenamefont {Bala{\v{z}}}, \citenamefont
  {Bassi}, \citenamefont {Bathe-Peters}, \citenamefont {Battelier},
  \citenamefont {Beli{\'c}}, \citenamefont {Bentine}, \citenamefont {Bernabeu},
  \citenamefont {Bertoldi}, \citenamefont {Bingham}, \citenamefont {Blas},
  \citenamefont {Bolpasi}, \citenamefont {Bongs}, \citenamefont {Bose},
  \citenamefont {Bouyer}, \citenamefont {Bowcock}, \citenamefont {Bowden},
  \citenamefont {Buchmueller}, \citenamefont {Burrage}, \citenamefont {Calmet},
  \citenamefont {Canuel}, \citenamefont {Caramete}, \citenamefont {Carroll},
  \citenamefont {Cella}, \citenamefont {Charmandaris}, \citenamefont
  {Chattopadhyay}, \citenamefont {Chen}, \citenamefont {Chiofalo},
  \citenamefont {Coleman}, \citenamefont {Cotter}, \citenamefont {Cui},
  \citenamefont {Derevianko}, \citenamefont {{De Roeck}}, \citenamefont
  {Djordjevic}, \citenamefont {Dornan}, \citenamefont {Doser}, \citenamefont
  {Drougkakis}, \citenamefont {Dunningham}, \citenamefont {Dutan},
  \citenamefont {Easo}, \citenamefont {Elertas}, \citenamefont {Ellis},
  \citenamefont {{El Sawy}}, \citenamefont {Fassi}, \citenamefont {Felea},
  \citenamefont {Feng}, \citenamefont {Flack}, \citenamefont {Foot},
  \citenamefont {Fuentes}, \citenamefont {Gaaloul}, \citenamefont {Gauguet},
  \citenamefont {Geiger}, \citenamefont {Gibson}, \citenamefont {Giudice},
  \citenamefont {Goldwin}, \citenamefont {Grachov}, \citenamefont {Graham},
  \citenamefont {Grasso}, \citenamefont {{van der Grinten}}, \citenamefont
  {G{\"u}ndogan}, \citenamefont {Haehnelt}, \citenamefont {Harte},
  \citenamefont {Hees}, \citenamefont {Hobson}, \citenamefont {Hogan},
  \citenamefont {Holst}, \citenamefont {Holynski}, \citenamefont {Kasevich},
  \citenamefont {Kavanagh}, \citenamefont {{von Klitzing}}, \citenamefont
  {Kovachy}, \citenamefont {Krikler}, \citenamefont {Krutzik}, \citenamefont
  {Lewicki}, \citenamefont {Lien}, \citenamefont {Liu}, \citenamefont
  {Luciano}, \citenamefont {Magnon}, \citenamefont {Mahmoud}, \citenamefont
  {Malik}, \citenamefont {McCabe}, \citenamefont {Mitchell}, \citenamefont
  {Pahl}, \citenamefont {Pal}, \citenamefont {Pandey}, \citenamefont
  {Papazoglou}, \citenamefont {Paternostro}, \citenamefont {Penning},
  \citenamefont {Peters}, \citenamefont {Prevedelli}, \citenamefont
  {Puthiya-Veettil}, \citenamefont {Quenby}, \citenamefont {Rasel},
  \citenamefont {Ravenhall}, \citenamefont {Ringwood}, \citenamefont {Roura},
  \citenamefont {Sabulsky}, \citenamefont {Sameed}, \citenamefont {Sauer},
  \citenamefont {Sch{\"a}ffer}, \citenamefont {Schiller}, \citenamefont
  {Schkolnik}, \citenamefont {Schlippert}, \citenamefont {Schubert},
  \citenamefont {Sfar}, \citenamefont {Shayeghi}, \citenamefont {Shipsey},
  \citenamefont {Signorini}, \citenamefont {Singh}, \citenamefont
  {Soares-Santos}, \citenamefont {Sorrentino}, \citenamefont {Sumner},
  \citenamefont {Tassis}, \citenamefont {Tentindo}, \citenamefont {Tino},
  \citenamefont {Tinsley}, \citenamefont {Unwin}, \citenamefont {Valenzuela},
  \citenamefont {Vasilakis}, \citenamefont {Vaskonen}, \citenamefont {Vogt},
  \citenamefont {Webber-Date}, \citenamefont {Wenzlawski}, \citenamefont
  {Windpassinger}, \citenamefont {Woltmann}, \citenamefont {Yazgan},
  \citenamefont {Zhan}, \citenamefont {Zou},\ and\ \citenamefont
  {Zupan}}]{ElNeaj2020EPJQ}%
  \BibitemOpen
  \bibfield  {author} {\bibinfo {author} {\bibfnamefont {Y.~A.}\ \bibnamefont
  {El-Neaj}}, \bibinfo {author} {\bibfnamefont {C.}~\bibnamefont {Alpigiani}},
  \bibinfo {author} {\bibfnamefont {S.}~\bibnamefont {Amairi-Pyka}}, \bibinfo
  {author} {\bibfnamefont {H.}~\bibnamefont {Ara{\'u}jo}}, \bibinfo {author}
  {\bibfnamefont {A.}~\bibnamefont {Bala{\v{z}}}}, \bibinfo {author}
  {\bibfnamefont {A.}~\bibnamefont {Bassi}}, \bibinfo {author} {\bibfnamefont
  {L.}~\bibnamefont {Bathe-Peters}}, \bibinfo {author} {\bibfnamefont
  {B.}~\bibnamefont {Battelier}}, \bibinfo {author} {\bibfnamefont
  {A.}~\bibnamefont {Beli{\'c}}}, \bibinfo {author} {\bibfnamefont
  {E.}~\bibnamefont {Bentine}}, \bibinfo {author} {\bibfnamefont
  {J.}~\bibnamefont {Bernabeu}}, \bibinfo {author} {\bibfnamefont
  {A.}~\bibnamefont {Bertoldi}}, \bibinfo {author} {\bibfnamefont
  {R.}~\bibnamefont {Bingham}}, \bibinfo {author} {\bibfnamefont
  {D.}~\bibnamefont {Blas}}, \bibinfo {author} {\bibfnamefont {V.}~\bibnamefont
  {Bolpasi}}, \bibinfo {author} {\bibfnamefont {K.}~\bibnamefont {Bongs}},
  \bibinfo {author} {\bibfnamefont {S.}~\bibnamefont {Bose}}, \bibinfo {author}
  {\bibfnamefont {P.}~\bibnamefont {Bouyer}}, \bibinfo {author} {\bibfnamefont
  {T.}~\bibnamefont {Bowcock}}, \bibinfo {author} {\bibfnamefont
  {W.}~\bibnamefont {Bowden}}, \bibinfo {author} {\bibfnamefont
  {O.}~\bibnamefont {Buchmueller}}, \bibinfo {author} {\bibfnamefont
  {C.}~\bibnamefont {Burrage}}, \bibinfo {author} {\bibfnamefont
  {X.}~\bibnamefont {Calmet}}, \bibinfo {author} {\bibfnamefont
  {B.}~\bibnamefont {Canuel}}, \bibinfo {author} {\bibfnamefont {L.-I.}\
  \bibnamefont {Caramete}}, \bibinfo {author} {\bibfnamefont {A.}~\bibnamefont
  {Carroll}}, \bibinfo {author} {\bibfnamefont {G.}~\bibnamefont {Cella}},
  \bibinfo {author} {\bibfnamefont {V.}~\bibnamefont {Charmandaris}}, \bibinfo
  {author} {\bibfnamefont {S.}~\bibnamefont {Chattopadhyay}}, \bibinfo {author}
  {\bibfnamefont {X.}~\bibnamefont {Chen}}, \bibinfo {author} {\bibfnamefont
  {M.~L.}\ \bibnamefont {Chiofalo}}, \bibinfo {author} {\bibfnamefont
  {J.}~\bibnamefont {Coleman}}, \bibinfo {author} {\bibfnamefont
  {J.}~\bibnamefont {Cotter}}, \bibinfo {author} {\bibfnamefont
  {Y.}~\bibnamefont {Cui}}, \bibinfo {author} {\bibfnamefont {A.}~\bibnamefont
  {Derevianko}}, \bibinfo {author} {\bibfnamefont {A.}~\bibnamefont {{De
  Roeck}}}, \bibinfo {author} {\bibfnamefont {G.~S.}\ \bibnamefont
  {Djordjevic}}, \bibinfo {author} {\bibfnamefont {P.}~\bibnamefont {Dornan}},
  \bibinfo {author} {\bibfnamefont {M.}~\bibnamefont {Doser}}, \bibinfo
  {author} {\bibfnamefont {I.}~\bibnamefont {Drougkakis}}, \bibinfo {author}
  {\bibfnamefont {J.}~\bibnamefont {Dunningham}}, \bibinfo {author}
  {\bibfnamefont {I.}~\bibnamefont {Dutan}}, \bibinfo {author} {\bibfnamefont
  {S.}~\bibnamefont {Easo}}, \bibinfo {author} {\bibfnamefont {G.}~\bibnamefont
  {Elertas}}, \bibinfo {author} {\bibfnamefont {J.}~\bibnamefont {Ellis}},
  \bibinfo {author} {\bibfnamefont {M.}~\bibnamefont {{El Sawy}}}, \bibinfo
  {author} {\bibfnamefont {F.}~\bibnamefont {Fassi}}, \bibinfo {author}
  {\bibfnamefont {D.}~\bibnamefont {Felea}}, \bibinfo {author} {\bibfnamefont
  {C.-H.}\ \bibnamefont {Feng}}, \bibinfo {author} {\bibfnamefont
  {R.}~\bibnamefont {Flack}}, \bibinfo {author} {\bibfnamefont
  {C.}~\bibnamefont {Foot}}, \bibinfo {author} {\bibfnamefont {I.}~\bibnamefont
  {Fuentes}}, \bibinfo {author} {\bibfnamefont {N.}~\bibnamefont {Gaaloul}},
  \bibinfo {author} {\bibfnamefont {A.}~\bibnamefont {Gauguet}}, \bibinfo
  {author} {\bibfnamefont {R.}~\bibnamefont {Geiger}}, \bibinfo {author}
  {\bibfnamefont {V.}~\bibnamefont {Gibson}}, \bibinfo {author} {\bibfnamefont
  {G.}~\bibnamefont {Giudice}}, \bibinfo {author} {\bibfnamefont
  {J.}~\bibnamefont {Goldwin}}, \bibinfo {author} {\bibfnamefont
  {O.}~\bibnamefont {Grachov}}, \bibinfo {author} {\bibfnamefont {P.~W.}\
  \bibnamefont {Graham}}, \bibinfo {author} {\bibfnamefont {D.}~\bibnamefont
  {Grasso}}, \bibinfo {author} {\bibfnamefont {M.}~\bibnamefont {{van der
  Grinten}}}, \bibinfo {author} {\bibfnamefont {M.}~\bibnamefont
  {G{\"u}ndogan}}, \bibinfo {author} {\bibfnamefont {M.~G.}\ \bibnamefont
  {Haehnelt}}, \bibinfo {author} {\bibfnamefont {T.}~\bibnamefont {Harte}},
  \bibinfo {author} {\bibfnamefont {A.}~\bibnamefont {Hees}}, \bibinfo {author}
  {\bibfnamefont {R.}~\bibnamefont {Hobson}}, \bibinfo {author} {\bibfnamefont
  {J.}~\bibnamefont {Hogan}}, \bibinfo {author} {\bibfnamefont
  {B.}~\bibnamefont {Holst}}, \bibinfo {author} {\bibfnamefont
  {M.}~\bibnamefont {Holynski}}, \bibinfo {author} {\bibfnamefont
  {M.}~\bibnamefont {Kasevich}}, \bibinfo {author} {\bibfnamefont {B.~J.}\
  \bibnamefont {Kavanagh}}, \bibinfo {author} {\bibfnamefont {W.}~\bibnamefont
  {{von Klitzing}}}, \bibinfo {author} {\bibfnamefont {T.}~\bibnamefont
  {Kovachy}}, \bibinfo {author} {\bibfnamefont {B.}~\bibnamefont {Krikler}},
  \bibinfo {author} {\bibfnamefont {M.}~\bibnamefont {Krutzik}}, \bibinfo
  {author} {\bibfnamefont {M.}~\bibnamefont {Lewicki}}, \bibinfo {author}
  {\bibfnamefont {Y.-H.}\ \bibnamefont {Lien}}, \bibinfo {author}
  {\bibfnamefont {M.}~\bibnamefont {Liu}}, \bibinfo {author} {\bibfnamefont
  {G.~G.}\ \bibnamefont {Luciano}}, \bibinfo {author} {\bibfnamefont
  {A.}~\bibnamefont {Magnon}}, \bibinfo {author} {\bibfnamefont {M.~A.}\
  \bibnamefont {Mahmoud}}, \bibinfo {author} {\bibfnamefont {S.}~\bibnamefont
  {Malik}}, \bibinfo {author} {\bibfnamefont {C.}~\bibnamefont {McCabe}},
  \bibinfo {author} {\bibfnamefont {J.}~\bibnamefont {Mitchell}}, \bibinfo
  {author} {\bibfnamefont {J.}~\bibnamefont {Pahl}}, \bibinfo {author}
  {\bibfnamefont {D.}~\bibnamefont {Pal}}, \bibinfo {author} {\bibfnamefont
  {S.}~\bibnamefont {Pandey}}, \bibinfo {author} {\bibfnamefont
  {D.}~\bibnamefont {Papazoglou}}, \bibinfo {author} {\bibfnamefont
  {M.}~\bibnamefont {Paternostro}}, \bibinfo {author} {\bibfnamefont
  {B.}~\bibnamefont {Penning}}, \bibinfo {author} {\bibfnamefont
  {A.}~\bibnamefont {Peters}}, \bibinfo {author} {\bibfnamefont
  {M.}~\bibnamefont {Prevedelli}}, \bibinfo {author} {\bibfnamefont
  {V.}~\bibnamefont {Puthiya-Veettil}}, \bibinfo {author} {\bibfnamefont
  {J.}~\bibnamefont {Quenby}}, \bibinfo {author} {\bibfnamefont
  {E.}~\bibnamefont {Rasel}}, \bibinfo {author} {\bibfnamefont
  {S.}~\bibnamefont {Ravenhall}}, \bibinfo {author} {\bibfnamefont
  {J.}~\bibnamefont {Ringwood}}, \bibinfo {author} {\bibfnamefont
  {A.}~\bibnamefont {Roura}}, \bibinfo {author} {\bibfnamefont
  {D.}~\bibnamefont {Sabulsky}}, \bibinfo {author} {\bibfnamefont
  {M.}~\bibnamefont {Sameed}}, \bibinfo {author} {\bibfnamefont
  {B.}~\bibnamefont {Sauer}}, \bibinfo {author} {\bibfnamefont {S.~A.}\
  \bibnamefont {Sch{\"a}ffer}}, \bibinfo {author} {\bibfnamefont
  {S.}~\bibnamefont {Schiller}}, \bibinfo {author} {\bibfnamefont
  {V.}~\bibnamefont {Schkolnik}}, \bibinfo {author} {\bibfnamefont
  {D.}~\bibnamefont {Schlippert}}, \bibinfo {author} {\bibfnamefont
  {C.}~\bibnamefont {Schubert}}, \bibinfo {author} {\bibfnamefont {H.~R.}\
  \bibnamefont {Sfar}}, \bibinfo {author} {\bibfnamefont {A.}~\bibnamefont
  {Shayeghi}}, \bibinfo {author} {\bibfnamefont {I.}~\bibnamefont {Shipsey}},
  \bibinfo {author} {\bibfnamefont {C.}~\bibnamefont {Signorini}}, \bibinfo
  {author} {\bibfnamefont {Y.}~\bibnamefont {Singh}}, \bibinfo {author}
  {\bibfnamefont {M.}~\bibnamefont {Soares-Santos}}, \bibinfo {author}
  {\bibfnamefont {F.}~\bibnamefont {Sorrentino}}, \bibinfo {author}
  {\bibfnamefont {T.}~\bibnamefont {Sumner}}, \bibinfo {author} {\bibfnamefont
  {K.}~\bibnamefont {Tassis}}, \bibinfo {author} {\bibfnamefont
  {S.}~\bibnamefont {Tentindo}}, \bibinfo {author} {\bibfnamefont {G.~M.}\
  \bibnamefont {Tino}}, \bibinfo {author} {\bibfnamefont {J.~N.}\ \bibnamefont
  {Tinsley}}, \bibinfo {author} {\bibfnamefont {J.}~\bibnamefont {Unwin}},
  \bibinfo {author} {\bibfnamefont {T.}~\bibnamefont {Valenzuela}}, \bibinfo
  {author} {\bibfnamefont {G.}~\bibnamefont {Vasilakis}}, \bibinfo {author}
  {\bibfnamefont {V.}~\bibnamefont {Vaskonen}}, \bibinfo {author}
  {\bibfnamefont {C.}~\bibnamefont {Vogt}}, \bibinfo {author} {\bibfnamefont
  {A.}~\bibnamefont {Webber-Date}}, \bibinfo {author} {\bibfnamefont
  {A.}~\bibnamefont {Wenzlawski}}, \bibinfo {author} {\bibfnamefont
  {P.}~\bibnamefont {Windpassinger}}, \bibinfo {author} {\bibfnamefont
  {M.}~\bibnamefont {Woltmann}}, \bibinfo {author} {\bibfnamefont
  {E.}~\bibnamefont {Yazgan}}, \bibinfo {author} {\bibfnamefont {M.-S.}\
  \bibnamefont {Zhan}}, \bibinfo {author} {\bibfnamefont {X.}~\bibnamefont
  {Zou}}, \ and\ \bibinfo {author} {\bibfnamefont {J.}~\bibnamefont {Zupan}},\
  }\href {\doibase 10.1140/epjqt/s40507-020-0080-0} {\bibfield  {journal}
  {\bibinfo  {journal} {EPJ Quantum Technology}\ }\textbf {\bibinfo {volume}
  {7}} (\bibinfo {year} {2020}),\ 10.1140/epjqt/s40507-020-0080-0}\BibitemShut
  {NoStop}%
\bibitem [{\citenamefont {Du}\ \emph {et~al.}(2022)\citenamefont {Du},
  \citenamefont {Murgui}, \citenamefont {Pardo}, \citenamefont {Wang},\ and\
  \citenamefont {Zurek}}]{Du2022PRD}%
  \BibitemOpen
  \bibfield  {author} {\bibinfo {author} {\bibfnamefont {Y.}~\bibnamefont
  {Du}}, \bibinfo {author} {\bibfnamefont {C.}~\bibnamefont {Murgui}}, \bibinfo
  {author} {\bibfnamefont {K.}~\bibnamefont {Pardo}}, \bibinfo {author}
  {\bibfnamefont {Y.}~\bibnamefont {Wang}}, \ and\ \bibinfo {author}
  {\bibfnamefont {K.~M.}\ \bibnamefont {Zurek}},\ }\href {\doibase
  10.1103/PhysRevD.106.095041} {\bibfield  {journal} {\bibinfo  {journal}
  {Phys. Rev. D}\ }\textbf {\bibinfo {volume} {106}},\ \bibinfo {pages}
  {095041} (\bibinfo {year} {2022})}\BibitemShut {NoStop}%
\bibitem [{\citenamefont {Badurina}\ \emph {et~al.}(2023)\citenamefont
  {Badurina}, \citenamefont {Gibson}, \citenamefont {McCabe},\ and\
  \citenamefont {Mitchell}}]{Badurina2023PRD}%
  \BibitemOpen
  \bibfield  {author} {\bibinfo {author} {\bibfnamefont {L.}~\bibnamefont
  {Badurina}}, \bibinfo {author} {\bibfnamefont {V.}~\bibnamefont {Gibson}},
  \bibinfo {author} {\bibfnamefont {C.}~\bibnamefont {McCabe}}, \ and\ \bibinfo
  {author} {\bibfnamefont {J.}~\bibnamefont {Mitchell}},\ }\href {\doibase
  10.1103/PhysRevD.107.055002} {\bibfield  {journal} {\bibinfo  {journal}
  {Phys. Rev. D}\ }\textbf {\bibinfo {volume} {107}},\ \bibinfo {pages}
  {055002} (\bibinfo {year} {2023})}\BibitemShut {NoStop}%
\bibitem [{\citenamefont {Hogan}\ \emph {et~al.}(2011)\citenamefont {Hogan},
  \citenamefont {Johnson}, \citenamefont {Dickerson}, \citenamefont {Kovachy},
  \citenamefont {Sugarbaker}, \citenamefont {Chiow}, \citenamefont {Graham},
  \citenamefont {Kasevich}, \citenamefont {Saif}, \citenamefont {Rajendran},
  \citenamefont {Bouyer}, \citenamefont {Seery}, \citenamefont {Feinberg},\
  and\ \citenamefont {Keski-Kuha}}]{Hogan2011GRG}%
  \BibitemOpen
  \bibfield  {author} {\bibinfo {author} {\bibfnamefont {J.~M.}\ \bibnamefont
  {Hogan}}, \bibinfo {author} {\bibfnamefont {D.~M.~S.}\ \bibnamefont
  {Johnson}}, \bibinfo {author} {\bibfnamefont {S.}~\bibnamefont {Dickerson}},
  \bibinfo {author} {\bibfnamefont {T.}~\bibnamefont {Kovachy}}, \bibinfo
  {author} {\bibfnamefont {A.}~\bibnamefont {Sugarbaker}}, \bibinfo {author}
  {\bibfnamefont {S.-w.}\ \bibnamefont {Chiow}}, \bibinfo {author}
  {\bibfnamefont {P.~W.}\ \bibnamefont {Graham}}, \bibinfo {author}
  {\bibfnamefont {M.~A.}\ \bibnamefont {Kasevich}}, \bibinfo {author}
  {\bibfnamefont {B.}~\bibnamefont {Saif}}, \bibinfo {author} {\bibfnamefont
  {S.}~\bibnamefont {Rajendran}}, \bibinfo {author} {\bibfnamefont
  {P.}~\bibnamefont {Bouyer}}, \bibinfo {author} {\bibfnamefont {B.~D.}\
  \bibnamefont {Seery}}, \bibinfo {author} {\bibfnamefont {L.}~\bibnamefont
  {Feinberg}}, \ and\ \bibinfo {author} {\bibfnamefont {R.}~\bibnamefont
  {Keski-Kuha}},\ }\href {\doibase 10.1007/s10714-011-1182-x} {\bibfield
  {journal} {\bibinfo  {journal} {General Relativity and Gravitation}\ }\textbf
  {\bibinfo {volume} {43}},\ \bibinfo {pages} {1953} (\bibinfo {year}
  {2011})}\BibitemShut {NoStop}%
\bibitem [{\citenamefont {Canuel}\ \emph {et~al.}(2018)\citenamefont {Canuel},
  \citenamefont {Bertoldi}, \citenamefont {Amand}, \citenamefont {{Di Pozzo
  Borgo}}, \citenamefont {Chantrait}, \citenamefont {Danquigny}, \citenamefont
  {{Dovale {\'A}lvarez}}, \citenamefont {Fang}, \citenamefont {Freise},
  \citenamefont {Geiger}, \citenamefont {Gillot}, \citenamefont {Henry},
  \citenamefont {Hinderer}, \citenamefont {Holleville}, \citenamefont {Junca},
  \citenamefont {Lef{\`e}vre}, \citenamefont {Merzougui}, \citenamefont
  {Mielec}, \citenamefont {Monfret}, \citenamefont {Pelisson}, \citenamefont
  {Prevedelli}, \citenamefont {Reynaud}, \citenamefont {Riou}, \citenamefont
  {Rogister}, \citenamefont {Rosat}, \citenamefont {Cormier}, \citenamefont
  {Landragin}, \citenamefont {Chaibi}, \citenamefont {Gaffet},\ and\
  \citenamefont {Bouyer}}]{Canuel2018SciRep}%
  \BibitemOpen
  \bibfield  {author} {\bibinfo {author} {\bibfnamefont {B.}~\bibnamefont
  {Canuel}}, \bibinfo {author} {\bibfnamefont {A.}~\bibnamefont {Bertoldi}},
  \bibinfo {author} {\bibfnamefont {L.}~\bibnamefont {Amand}}, \bibinfo
  {author} {\bibfnamefont {E.}~\bibnamefont {{Di Pozzo Borgo}}}, \bibinfo
  {author} {\bibfnamefont {T.}~\bibnamefont {Chantrait}}, \bibinfo {author}
  {\bibfnamefont {C.}~\bibnamefont {Danquigny}}, \bibinfo {author}
  {\bibfnamefont {M.}~\bibnamefont {{Dovale {\'A}lvarez}}}, \bibinfo {author}
  {\bibfnamefont {B.}~\bibnamefont {Fang}}, \bibinfo {author} {\bibfnamefont
  {A.}~\bibnamefont {Freise}}, \bibinfo {author} {\bibfnamefont
  {R.}~\bibnamefont {Geiger}}, \bibinfo {author} {\bibfnamefont
  {J.}~\bibnamefont {Gillot}}, \bibinfo {author} {\bibfnamefont
  {S.}~\bibnamefont {Henry}}, \bibinfo {author} {\bibfnamefont
  {J.}~\bibnamefont {Hinderer}}, \bibinfo {author} {\bibfnamefont
  {D.}~\bibnamefont {Holleville}}, \bibinfo {author} {\bibfnamefont
  {J.}~\bibnamefont {Junca}}, \bibinfo {author} {\bibfnamefont
  {G.}~\bibnamefont {Lef{\`e}vre}}, \bibinfo {author} {\bibfnamefont
  {M.}~\bibnamefont {Merzougui}}, \bibinfo {author} {\bibfnamefont
  {N.}~\bibnamefont {Mielec}}, \bibinfo {author} {\bibfnamefont
  {T.}~\bibnamefont {Monfret}}, \bibinfo {author} {\bibfnamefont
  {S.}~\bibnamefont {Pelisson}}, \bibinfo {author} {\bibfnamefont
  {M.}~\bibnamefont {Prevedelli}}, \bibinfo {author} {\bibfnamefont
  {S.}~\bibnamefont {Reynaud}}, \bibinfo {author} {\bibfnamefont
  {I.}~\bibnamefont {Riou}}, \bibinfo {author} {\bibfnamefont {Y.}~\bibnamefont
  {Rogister}}, \bibinfo {author} {\bibfnamefont {S.}~\bibnamefont {Rosat}},
  \bibinfo {author} {\bibfnamefont {E.}~\bibnamefont {Cormier}}, \bibinfo
  {author} {\bibfnamefont {A.}~\bibnamefont {Landragin}}, \bibinfo {author}
  {\bibfnamefont {W.}~\bibnamefont {Chaibi}}, \bibinfo {author} {\bibfnamefont
  {S.}~\bibnamefont {Gaffet}}, \ and\ \bibinfo {author} {\bibfnamefont
  {P.}~\bibnamefont {Bouyer}},\ }\href {\doibase 10.1038/s41598-018-32165-z}
  {\bibfield  {journal} {\bibinfo  {journal} {Scientific reports}\ }\textbf
  {\bibinfo {volume} {8}},\ \bibinfo {pages} {14064} (\bibinfo {year}
  {2018})}\BibitemShut {NoStop}%
\bibitem [{\citenamefont {Zhan}\ \emph {et~al.}(2020)\citenamefont {Zhan},
  \citenamefont {Wang}, \citenamefont {Ni}, \citenamefont {Gao}, \citenamefont
  {Wang}, \citenamefont {He}, \citenamefont {Li}, \citenamefont {Zhou},
  \citenamefont {Chen}, \citenamefont {Zhong}, \citenamefont {Tang},
  \citenamefont {Yao}, \citenamefont {Zhu}, \citenamefont {Xiong},
  \citenamefont {Lu}, \citenamefont {Yu}, \citenamefont {Cheng}, \citenamefont
  {Liu}, \citenamefont {Liang}, \citenamefont {Xu}, \citenamefont {He},
  \citenamefont {Ke}, \citenamefont {Tan},\ and\ \citenamefont
  {Luo}}]{Zhan2019quq}%
  \BibitemOpen
  \bibfield  {author} {\bibinfo {author} {\bibfnamefont {M.-S.}\ \bibnamefont
  {Zhan}}, \bibinfo {author} {\bibfnamefont {J.}~\bibnamefont {Wang}}, \bibinfo
  {author} {\bibfnamefont {W.-T.}\ \bibnamefont {Ni}}, \bibinfo {author}
  {\bibfnamefont {D.-F.}\ \bibnamefont {Gao}}, \bibinfo {author} {\bibfnamefont
  {G.}~\bibnamefont {Wang}}, \bibinfo {author} {\bibfnamefont {L.-X.}\
  \bibnamefont {He}}, \bibinfo {author} {\bibfnamefont {R.-B.}\ \bibnamefont
  {Li}}, \bibinfo {author} {\bibfnamefont {L.}~\bibnamefont {Zhou}}, \bibinfo
  {author} {\bibfnamefont {X.}~\bibnamefont {Chen}}, \bibinfo {author}
  {\bibfnamefont {J.-Q.}\ \bibnamefont {Zhong}}, \bibinfo {author}
  {\bibfnamefont {B.}~\bibnamefont {Tang}}, \bibinfo {author} {\bibfnamefont
  {Z.-W.}\ \bibnamefont {Yao}}, \bibinfo {author} {\bibfnamefont
  {L.}~\bibnamefont {Zhu}}, \bibinfo {author} {\bibfnamefont {Z.-Y.}\
  \bibnamefont {Xiong}}, \bibinfo {author} {\bibfnamefont {S.-B.}\ \bibnamefont
  {Lu}}, \bibinfo {author} {\bibfnamefont {G.-H.}\ \bibnamefont {Yu}}, \bibinfo
  {author} {\bibfnamefont {Q.-F.}\ \bibnamefont {Cheng}}, \bibinfo {author}
  {\bibfnamefont {M.}~\bibnamefont {Liu}}, \bibinfo {author} {\bibfnamefont
  {Y.-R.}\ \bibnamefont {Liang}}, \bibinfo {author} {\bibfnamefont
  {P.}~\bibnamefont {Xu}}, \bibinfo {author} {\bibfnamefont {X.-D.}\
  \bibnamefont {He}}, \bibinfo {author} {\bibfnamefont {M.}~\bibnamefont {Ke}},
  \bibinfo {author} {\bibfnamefont {Z.}~\bibnamefont {Tan}}, \ and\ \bibinfo
  {author} {\bibfnamefont {J.}~\bibnamefont {Luo}},\ }\href {\doibase
  10.1142/S0218271819400054} {\bibfield  {journal} {\bibinfo  {journal}
  {International Journal of Modern Physics D}\ }\textbf {\bibinfo {volume}
  {29}},\ \bibinfo {pages} {1940005} (\bibinfo {year} {2020})}\BibitemShut
  {NoStop}%
\bibitem [{\citenamefont {Schubert}\ \emph {et~al.}(2019)\citenamefont
  {Schubert}, \citenamefont {Schlippert}, \citenamefont {Abend}, \citenamefont
  {Giese}, \citenamefont {Roura}, \citenamefont {Schleich}, \citenamefont
  {Ertmer},\ and\ \citenamefont {Rasel}}]{schubert_scalable_2019}%
  \BibitemOpen
  \bibfield  {author} {\bibinfo {author} {\bibfnamefont {C.}~\bibnamefont
  {Schubert}}, \bibinfo {author} {\bibfnamefont {D.}~\bibnamefont
  {Schlippert}}, \bibinfo {author} {\bibfnamefont {S.}~\bibnamefont {Abend}},
  \bibinfo {author} {\bibfnamefont {E.}~\bibnamefont {Giese}}, \bibinfo
  {author} {\bibfnamefont {A.}~\bibnamefont {Roura}}, \bibinfo {author}
  {\bibfnamefont {W.~P.}\ \bibnamefont {Schleich}}, \bibinfo {author}
  {\bibfnamefont {W.}~\bibnamefont {Ertmer}}, \ and\ \bibinfo {author}
  {\bibfnamefont {E.~M.}\ \bibnamefont {Rasel}},\ }\href {\doibase
  10.48550/arXiv.1909.01951} {\enquote {\bibinfo {title} {Scalable, symmetric
  atom interferometer for infrasound gravitational wave detection},}\ }
  (\bibinfo {year} {2019}),\ \bibinfo {note} {arXiv:1909.01951
  [quant-ph]}\BibitemShut {NoStop}%
\bibitem [{\citenamefont {Canuel}\ \emph {et~al.}(2020)\citenamefont {Canuel},
  \citenamefont {Abend}, \citenamefont {Amaro-Seoane}, \citenamefont
  {Badaracco}, \citenamefont {Beaufils}, \citenamefont {Bertoldi},
  \citenamefont {Bongs}, \citenamefont {Bouyer}, \citenamefont {Braxmaier},
  \citenamefont {Chaibi}, \citenamefont {Christensen}, \citenamefont {Fitzek},
  \citenamefont {Flouris}, \citenamefont {Gaaloul}, \citenamefont {Gaffet},
  \citenamefont {Alzar}, \citenamefont {Geiger}, \citenamefont
  {Guellati-Khelifa}, \citenamefont {Hammerer}, \citenamefont {Harms},
  \citenamefont {Hinderer}, \citenamefont {Holynski}, \citenamefont {Junca},
  \citenamefont {Katsanevas}, \citenamefont {Klempt}, \citenamefont
  {Kozanitis}, \citenamefont {Krutzik}, \citenamefont {Landragin},
  \citenamefont {Roche}, \citenamefont {Leykauf}, \citenamefont {Lien},
  \citenamefont {Loriani}, \citenamefont {Merlet}, \citenamefont {Merzougui},
  \citenamefont {Nofrarias}, \citenamefont {Papadakos}, \citenamefont {dos
  Santos}, \citenamefont {Peters}, \citenamefont {Plexousakis}, \citenamefont
  {Prevedelli}, \citenamefont {Rasel}, \citenamefont {Rogister}, \citenamefont
  {Rosat}, \citenamefont {Roura}, \citenamefont {Sabulsky}, \citenamefont
  {Schkolnik}, \citenamefont {Schlippert}, \citenamefont {Schubert},
  \citenamefont {Sidorenkov}, \citenamefont {Siemß}, \citenamefont {Sopuerta},
  \citenamefont {Sorrentino}, \citenamefont {Struckmann}, \citenamefont {Tino},
  \citenamefont {Tsagkatakis}, \citenamefont {Viceré}, \citenamefont {von
  Klitzing}, \citenamefont {Woerner},\ and\ \citenamefont
  {Zou}}]{Canuel2020CQG}%
  \BibitemOpen
  \bibfield  {author} {\bibinfo {author} {\bibfnamefont {B.}~\bibnamefont
  {Canuel}}, \bibinfo {author} {\bibfnamefont {S.}~\bibnamefont {Abend}},
  \bibinfo {author} {\bibfnamefont {P.}~\bibnamefont {Amaro-Seoane}}, \bibinfo
  {author} {\bibfnamefont {F.}~\bibnamefont {Badaracco}}, \bibinfo {author}
  {\bibfnamefont {Q.}~\bibnamefont {Beaufils}}, \bibinfo {author}
  {\bibfnamefont {A.}~\bibnamefont {Bertoldi}}, \bibinfo {author}
  {\bibfnamefont {K.}~\bibnamefont {Bongs}}, \bibinfo {author} {\bibfnamefont
  {P.}~\bibnamefont {Bouyer}}, \bibinfo {author} {\bibfnamefont
  {C.}~\bibnamefont {Braxmaier}}, \bibinfo {author} {\bibfnamefont
  {W.}~\bibnamefont {Chaibi}}, \bibinfo {author} {\bibfnamefont
  {N.}~\bibnamefont {Christensen}}, \bibinfo {author} {\bibfnamefont
  {F.}~\bibnamefont {Fitzek}}, \bibinfo {author} {\bibfnamefont
  {G.}~\bibnamefont {Flouris}}, \bibinfo {author} {\bibfnamefont
  {N.}~\bibnamefont {Gaaloul}}, \bibinfo {author} {\bibfnamefont
  {S.}~\bibnamefont {Gaffet}}, \bibinfo {author} {\bibfnamefont {C.~L.~G.}\
  \bibnamefont {Alzar}}, \bibinfo {author} {\bibfnamefont {R.}~\bibnamefont
  {Geiger}}, \bibinfo {author} {\bibfnamefont {S.}~\bibnamefont
  {Guellati-Khelifa}}, \bibinfo {author} {\bibfnamefont {K.}~\bibnamefont
  {Hammerer}}, \bibinfo {author} {\bibfnamefont {J.}~\bibnamefont {Harms}},
  \bibinfo {author} {\bibfnamefont {J.}~\bibnamefont {Hinderer}}, \bibinfo
  {author} {\bibfnamefont {M.}~\bibnamefont {Holynski}}, \bibinfo {author}
  {\bibfnamefont {J.}~\bibnamefont {Junca}}, \bibinfo {author} {\bibfnamefont
  {S.}~\bibnamefont {Katsanevas}}, \bibinfo {author} {\bibfnamefont
  {C.}~\bibnamefont {Klempt}}, \bibinfo {author} {\bibfnamefont
  {C.}~\bibnamefont {Kozanitis}}, \bibinfo {author} {\bibfnamefont
  {M.}~\bibnamefont {Krutzik}}, \bibinfo {author} {\bibfnamefont
  {A.}~\bibnamefont {Landragin}}, \bibinfo {author} {\bibfnamefont {I.~L.}\
  \bibnamefont {Roche}}, \bibinfo {author} {\bibfnamefont {B.}~\bibnamefont
  {Leykauf}}, \bibinfo {author} {\bibfnamefont {Y.-H.}\ \bibnamefont {Lien}},
  \bibinfo {author} {\bibfnamefont {S.}~\bibnamefont {Loriani}}, \bibinfo
  {author} {\bibfnamefont {S.}~\bibnamefont {Merlet}}, \bibinfo {author}
  {\bibfnamefont {M.}~\bibnamefont {Merzougui}}, \bibinfo {author}
  {\bibfnamefont {M.}~\bibnamefont {Nofrarias}}, \bibinfo {author}
  {\bibfnamefont {P.}~\bibnamefont {Papadakos}}, \bibinfo {author}
  {\bibfnamefont {F.~P.}\ \bibnamefont {dos Santos}}, \bibinfo {author}
  {\bibfnamefont {A.}~\bibnamefont {Peters}}, \bibinfo {author} {\bibfnamefont
  {D.}~\bibnamefont {Plexousakis}}, \bibinfo {author} {\bibfnamefont
  {M.}~\bibnamefont {Prevedelli}}, \bibinfo {author} {\bibfnamefont {E.~M.}\
  \bibnamefont {Rasel}}, \bibinfo {author} {\bibfnamefont {Y.}~\bibnamefont
  {Rogister}}, \bibinfo {author} {\bibfnamefont {S.}~\bibnamefont {Rosat}},
  \bibinfo {author} {\bibfnamefont {A.}~\bibnamefont {Roura}}, \bibinfo
  {author} {\bibfnamefont {D.~O.}\ \bibnamefont {Sabulsky}}, \bibinfo {author}
  {\bibfnamefont {V.}~\bibnamefont {Schkolnik}}, \bibinfo {author}
  {\bibfnamefont {D.}~\bibnamefont {Schlippert}}, \bibinfo {author}
  {\bibfnamefont {C.}~\bibnamefont {Schubert}}, \bibinfo {author}
  {\bibfnamefont {L.}~\bibnamefont {Sidorenkov}}, \bibinfo {author}
  {\bibfnamefont {J.-N.}\ \bibnamefont {Siemß}}, \bibinfo {author}
  {\bibfnamefont {C.~F.}\ \bibnamefont {Sopuerta}}, \bibinfo {author}
  {\bibfnamefont {F.}~\bibnamefont {Sorrentino}}, \bibinfo {author}
  {\bibfnamefont {C.}~\bibnamefont {Struckmann}}, \bibinfo {author}
  {\bibfnamefont {G.~M.}\ \bibnamefont {Tino}}, \bibinfo {author}
  {\bibfnamefont {G.}~\bibnamefont {Tsagkatakis}}, \bibinfo {author}
  {\bibfnamefont {A.}~\bibnamefont {Viceré}}, \bibinfo {author} {\bibfnamefont
  {W.}~\bibnamefont {von Klitzing}}, \bibinfo {author} {\bibfnamefont
  {L.}~\bibnamefont {Woerner}}, \ and\ \bibinfo {author} {\bibfnamefont
  {X.}~\bibnamefont {Zou}},\ }\href {\doibase 10.1088/1361-6382/aba80e}
  {\bibfield  {journal} {\bibinfo  {journal} {Classical and Quantum Gravity}\
  }\textbf {\bibinfo {volume} {37}},\ \bibinfo {pages} {225017} (\bibinfo
  {year} {2020})}\BibitemShut {NoStop}%
\bibitem [{\citenamefont {Badurina}\ \emph {et~al.}(2020)\citenamefont
  {Badurina}, \citenamefont {Bentine}, \citenamefont {Blas}, \citenamefont
  {Bongs}, \citenamefont {Bortoletto}, \citenamefont {Bowcock}, \citenamefont
  {Bridges}, \citenamefont {Bowden}, \citenamefont {Buchmueller}, \citenamefont
  {Burrage}, \citenamefont {Coleman}, \citenamefont {Elertas}, \citenamefont
  {Ellis}, \citenamefont {Foot}, \citenamefont {Gibson}, \citenamefont
  {Haehnelt}, \citenamefont {Harte}, \citenamefont {Hedges}, \citenamefont
  {Hobson}, \citenamefont {Holynski}, \citenamefont {Jones}, \citenamefont
  {Langlois}, \citenamefont {Lellouch}, \citenamefont {Lewicki}, \citenamefont
  {Maiolino}, \citenamefont {Majewski}, \citenamefont {Malik}, \citenamefont
  {March-Russell}, \citenamefont {McCabe}, \citenamefont {Newbold},
  \citenamefont {Sauer}, \citenamefont {Schneider}, \citenamefont {Shipsey},
  \citenamefont {Singh}, \citenamefont {Uchida}, \citenamefont {Valenzuela},
  \citenamefont {van~der Grinten}, \citenamefont {Vaskonen}, \citenamefont
  {Vossebeld}, \citenamefont {Weatherill},\ and\ \citenamefont
  {Wilmut}}]{Badurina2020JCA}%
  \BibitemOpen
  \bibfield  {author} {\bibinfo {author} {\bibfnamefont {L.}~\bibnamefont
  {Badurina}}, \bibinfo {author} {\bibfnamefont {E.}~\bibnamefont {Bentine}},
  \bibinfo {author} {\bibfnamefont {D.}~\bibnamefont {Blas}}, \bibinfo {author}
  {\bibfnamefont {K.}~\bibnamefont {Bongs}}, \bibinfo {author} {\bibfnamefont
  {D.}~\bibnamefont {Bortoletto}}, \bibinfo {author} {\bibfnamefont
  {T.}~\bibnamefont {Bowcock}}, \bibinfo {author} {\bibfnamefont
  {K.}~\bibnamefont {Bridges}}, \bibinfo {author} {\bibfnamefont
  {W.}~\bibnamefont {Bowden}}, \bibinfo {author} {\bibfnamefont
  {O.}~\bibnamefont {Buchmueller}}, \bibinfo {author} {\bibfnamefont
  {C.}~\bibnamefont {Burrage}}, \bibinfo {author} {\bibfnamefont
  {J.}~\bibnamefont {Coleman}}, \bibinfo {author} {\bibfnamefont
  {G.}~\bibnamefont {Elertas}}, \bibinfo {author} {\bibfnamefont
  {J.}~\bibnamefont {Ellis}}, \bibinfo {author} {\bibfnamefont
  {C.}~\bibnamefont {Foot}}, \bibinfo {author} {\bibfnamefont {V.}~\bibnamefont
  {Gibson}}, \bibinfo {author} {\bibfnamefont {M.}~\bibnamefont {Haehnelt}},
  \bibinfo {author} {\bibfnamefont {T.}~\bibnamefont {Harte}}, \bibinfo
  {author} {\bibfnamefont {S.}~\bibnamefont {Hedges}}, \bibinfo {author}
  {\bibfnamefont {R.}~\bibnamefont {Hobson}}, \bibinfo {author} {\bibfnamefont
  {M.}~\bibnamefont {Holynski}}, \bibinfo {author} {\bibfnamefont
  {T.}~\bibnamefont {Jones}}, \bibinfo {author} {\bibfnamefont
  {M.}~\bibnamefont {Langlois}}, \bibinfo {author} {\bibfnamefont
  {S.}~\bibnamefont {Lellouch}}, \bibinfo {author} {\bibfnamefont
  {M.}~\bibnamefont {Lewicki}}, \bibinfo {author} {\bibfnamefont
  {R.}~\bibnamefont {Maiolino}}, \bibinfo {author} {\bibfnamefont
  {P.}~\bibnamefont {Majewski}}, \bibinfo {author} {\bibfnamefont
  {S.}~\bibnamefont {Malik}}, \bibinfo {author} {\bibfnamefont
  {J.}~\bibnamefont {March-Russell}}, \bibinfo {author} {\bibfnamefont
  {C.}~\bibnamefont {McCabe}}, \bibinfo {author} {\bibfnamefont
  {D.}~\bibnamefont {Newbold}}, \bibinfo {author} {\bibfnamefont
  {B.}~\bibnamefont {Sauer}}, \bibinfo {author} {\bibfnamefont
  {U.}~\bibnamefont {Schneider}}, \bibinfo {author} {\bibfnamefont
  {I.}~\bibnamefont {Shipsey}}, \bibinfo {author} {\bibfnamefont
  {Y.}~\bibnamefont {Singh}}, \bibinfo {author} {\bibfnamefont
  {M.}~\bibnamefont {Uchida}}, \bibinfo {author} {\bibfnamefont
  {T.}~\bibnamefont {Valenzuela}}, \bibinfo {author} {\bibfnamefont
  {M.}~\bibnamefont {van~der Grinten}}, \bibinfo {author} {\bibfnamefont
  {V.}~\bibnamefont {Vaskonen}}, \bibinfo {author} {\bibfnamefont
  {J.}~\bibnamefont {Vossebeld}}, \bibinfo {author} {\bibfnamefont
  {D.}~\bibnamefont {Weatherill}}, \ and\ \bibinfo {author} {\bibfnamefont
  {I.}~\bibnamefont {Wilmut}},\ }\href {\doibase 10.1088/1475-7516/2020/05/011}
  {\bibfield  {journal} {\bibinfo  {journal} {Journal of Cosmology and
  Astroparticle Physics}\ }\textbf {\bibinfo {volume} {2020}},\ \bibinfo
  {pages} {011} (\bibinfo {year} {2020})}\BibitemShut {NoStop}%
\bibitem [{\citenamefont {Anderson}\ \emph {et~al.}(1995)\citenamefont
  {Anderson}, \citenamefont {Ensher}, \citenamefont {Matthews}, \citenamefont
  {Wieman},\ and\ \citenamefont {Cornell}}]{Anderson95Science}%
  \BibitemOpen
  \bibfield  {author} {\bibinfo {author} {\bibfnamefont {M.}~\bibnamefont
  {Anderson}}, \bibinfo {author} {\bibfnamefont {J.}~\bibnamefont {Ensher}},
  \bibinfo {author} {\bibfnamefont {M.}~\bibnamefont {Matthews}}, \bibinfo
  {author} {\bibfnamefont {C.}~\bibnamefont {Wieman}}, \ and\ \bibinfo {author}
  {\bibfnamefont {E.}~\bibnamefont {Cornell}},\ }\href {\doibase
  10.1126/science.269.5221.198} {\bibfield  {journal} {\bibinfo  {journal}
  {Science}\ }\textbf {\bibinfo {volume} {269}},\ \bibinfo {pages} {198}
  (\bibinfo {year} {1995})}\BibitemShut {NoStop}%
\bibitem [{\citenamefont {Davis}\ \emph {et~al.}(1995)\citenamefont {Davis},
  \citenamefont {Mewes}, \citenamefont {Joffe}, \citenamefont {Andrews},\ and\
  \citenamefont {Ketterle}}]{Davis95PRL}%
  \BibitemOpen
  \bibfield  {author} {\bibinfo {author} {\bibfnamefont {K.~B.}\ \bibnamefont
  {Davis}}, \bibinfo {author} {\bibfnamefont {M.-O.}\ \bibnamefont {Mewes}},
  \bibinfo {author} {\bibfnamefont {M.~A.}\ \bibnamefont {Joffe}}, \bibinfo
  {author} {\bibfnamefont {M.~R.}\ \bibnamefont {Andrews}}, \ and\ \bibinfo
  {author} {\bibfnamefont {W.}~\bibnamefont {Ketterle}},\ }\href {\doibase
  10.1103/physrevlett.74.5202} {\bibfield  {journal} {\bibinfo  {journal}
  {Physical Review Letters}\ }\textbf {\bibinfo {volume} {74}},\ \bibinfo
  {pages} {5202} (\bibinfo {year} {1995})}\BibitemShut {NoStop}%
\bibitem [{\citenamefont {Schlippert}\ \emph {et~al.}(2021)\citenamefont
  {Schlippert}, \citenamefont {Meiners}, \citenamefont {Rengelink},
  \citenamefont {Schubert}, \citenamefont {Tell}, \citenamefont {Wodey},
  \citenamefont {Zipfel}, \citenamefont {Ertmer},\ and\ \citenamefont
  {Rasel}}]{Schlippert2020}%
  \BibitemOpen
  \bibfield  {author} {\bibinfo {author} {\bibfnamefont {D.}~\bibnamefont
  {Schlippert}}, \bibinfo {author} {\bibfnamefont {C.}~\bibnamefont {Meiners}},
  \bibinfo {author} {\bibfnamefont {R.}~\bibnamefont {Rengelink}}, \bibinfo
  {author} {\bibfnamefont {C.}~\bibnamefont {Schubert}}, \bibinfo {author}
  {\bibfnamefont {D.}~\bibnamefont {Tell}}, \bibinfo {author} {\bibfnamefont
  {{\'{E}}.}~\bibnamefont {Wodey}}, \bibinfo {author} {\bibfnamefont
  {K.}~\bibnamefont {Zipfel}}, \bibinfo {author} {\bibfnamefont
  {W.}~\bibnamefont {Ertmer}}, \ and\ \bibinfo {author} {\bibfnamefont
  {E.}~\bibnamefont {Rasel}},\ }in\ \href {\doibase 10.1142/9789811213984_0010}
  {\emph {\bibinfo {booktitle} {{CPT} and Lorentz Symmetry}}}\ (\bibinfo
  {publisher} {{WORLD} {SCIENTIFIC}},\ \bibinfo {year} {2021})\BibitemShut
  {NoStop}%
\bibitem [{\citenamefont {Hensel}\ \emph {et~al.}(2021)\citenamefont {Hensel},
  \citenamefont {Loriani}, \citenamefont {Schubert}, \citenamefont {Fitzek},
  \citenamefont {Abend}, \citenamefont {Ahlers}, \citenamefont {Siem{\ss}},
  \citenamefont {Hammerer}, \citenamefont {Rasel},\ and\ \citenamefont
  {Gaaloul}}]{Hensel2021}%
  \BibitemOpen
  \bibfield  {author} {\bibinfo {author} {\bibfnamefont {T.}~\bibnamefont
  {Hensel}}, \bibinfo {author} {\bibfnamefont {S.}~\bibnamefont {Loriani}},
  \bibinfo {author} {\bibfnamefont {C.}~\bibnamefont {Schubert}}, \bibinfo
  {author} {\bibfnamefont {F.}~\bibnamefont {Fitzek}}, \bibinfo {author}
  {\bibfnamefont {S.}~\bibnamefont {Abend}}, \bibinfo {author} {\bibfnamefont
  {H.}~\bibnamefont {Ahlers}}, \bibinfo {author} {\bibfnamefont {J.-N.}\
  \bibnamefont {Siem{\ss}}}, \bibinfo {author} {\bibfnamefont {K.}~\bibnamefont
  {Hammerer}}, \bibinfo {author} {\bibfnamefont {E.~M.}\ \bibnamefont {Rasel}},
  \ and\ \bibinfo {author} {\bibfnamefont {N.}~\bibnamefont {Gaaloul}},\ }\href
  {\doibase 10.1140/epjd/s10053-021-00069-9} {\bibfield  {journal} {\bibinfo
  {journal} {The European Physical Journal D}\ }\textbf {\bibinfo {volume}
  {75}},\ \bibinfo {pages} {108} (\bibinfo {year} {2021})}\BibitemShut
  {NoStop}%
\bibitem [{\citenamefont {Ammann}\ and\ \citenamefont
  {Christensen}(1997)}]{Ammann1997PRL}%
  \BibitemOpen
  \bibfield  {author} {\bibinfo {author} {\bibfnamefont {H.}~\bibnamefont
  {Ammann}}\ and\ \bibinfo {author} {\bibfnamefont {N.}~\bibnamefont
  {Christensen}},\ }\href {\doibase 10.1103/PhysRevLett.78.2088} {\bibfield
  {journal} {\bibinfo  {journal} {Phys. Rev. Lett.}\ }\textbf {\bibinfo
  {volume} {78}},\ \bibinfo {pages} {2088} (\bibinfo {year}
  {1997})}\BibitemShut {NoStop}%
\bibitem [{\citenamefont {Deppner}\ \emph {et~al.}(2021)\citenamefont
  {Deppner}, \citenamefont {Herr}, \citenamefont {Cornelius}, \citenamefont
  {Stromberger}, \citenamefont {Sternke}, \citenamefont {Grzeschik},
  \citenamefont {Grote}, \citenamefont {Rudolph}, \citenamefont {Herrmann},
  \citenamefont {Krutzik}, \citenamefont {Wenzlawski}, \citenamefont {Corgier},
  \citenamefont {Charron}, \citenamefont {Gu{\'e}ry-Odelin}, \citenamefont
  {Gaaloul}, \citenamefont {L{\"a}mmerzahl}, \citenamefont {Peters},
  \citenamefont {Windpassinger},\ and\ \citenamefont {Rasel}}]{Deppner21PRL}%
  \BibitemOpen
  \bibfield  {author} {\bibinfo {author} {\bibfnamefont {C.}~\bibnamefont
  {Deppner}}, \bibinfo {author} {\bibfnamefont {W.}~\bibnamefont {Herr}},
  \bibinfo {author} {\bibfnamefont {M.}~\bibnamefont {Cornelius}}, \bibinfo
  {author} {\bibfnamefont {P.}~\bibnamefont {Stromberger}}, \bibinfo {author}
  {\bibfnamefont {T.}~\bibnamefont {Sternke}}, \bibinfo {author} {\bibfnamefont
  {C.}~\bibnamefont {Grzeschik}}, \bibinfo {author} {\bibfnamefont
  {A.}~\bibnamefont {Grote}}, \bibinfo {author} {\bibfnamefont
  {J.}~\bibnamefont {Rudolph}}, \bibinfo {author} {\bibfnamefont
  {S.}~\bibnamefont {Herrmann}}, \bibinfo {author} {\bibfnamefont
  {M.}~\bibnamefont {Krutzik}}, \bibinfo {author} {\bibfnamefont
  {A.}~\bibnamefont {Wenzlawski}}, \bibinfo {author} {\bibfnamefont
  {R.}~\bibnamefont {Corgier}}, \bibinfo {author} {\bibfnamefont
  {E.}~\bibnamefont {Charron}}, \bibinfo {author} {\bibfnamefont
  {D.}~\bibnamefont {Gu{\'e}ry-Odelin}}, \bibinfo {author} {\bibfnamefont
  {N.}~\bibnamefont {Gaaloul}}, \bibinfo {author} {\bibfnamefont
  {C.}~\bibnamefont {L{\"a}mmerzahl}}, \bibinfo {author} {\bibfnamefont
  {A.}~\bibnamefont {Peters}}, \bibinfo {author} {\bibfnamefont
  {P.}~\bibnamefont {Windpassinger}}, \ and\ \bibinfo {author} {\bibfnamefont
  {E.~M.}\ \bibnamefont {Rasel}},\ }\href {\doibase
  10.1103/PhysRevLett.127.100401} {\bibfield  {journal} {\bibinfo  {journal}
  {Physical review letters}\ }\textbf {\bibinfo {volume} {127}},\ \bibinfo
  {pages} {100401} (\bibinfo {year} {2021})}\BibitemShut {NoStop}%
\bibitem [{\citenamefont {Rudolph}\ \emph {et~al.}(2015)\citenamefont
  {Rudolph}, \citenamefont {Herr}, \citenamefont {Grzeschik}, \citenamefont
  {Sternke}, \citenamefont {Grote}, \citenamefont {Popp}, \citenamefont
  {Becker}, \citenamefont {Müntinga}, \citenamefont {Ahlers}, \citenamefont
  {Peters}, \citenamefont {Lämmerzahl}, \citenamefont {Sengstock},
  \citenamefont {Gaaloul}, \citenamefont {Ertmer},\ and\ \citenamefont
  {Rasel}}]{Rudolph15NJP}%
  \BibitemOpen
  \bibfield  {author} {\bibinfo {author} {\bibfnamefont {J.}~\bibnamefont
  {Rudolph}}, \bibinfo {author} {\bibfnamefont {W.}~\bibnamefont {Herr}},
  \bibinfo {author} {\bibfnamefont {C.}~\bibnamefont {Grzeschik}}, \bibinfo
  {author} {\bibfnamefont {T.}~\bibnamefont {Sternke}}, \bibinfo {author}
  {\bibfnamefont {A.}~\bibnamefont {Grote}}, \bibinfo {author} {\bibfnamefont
  {M.}~\bibnamefont {Popp}}, \bibinfo {author} {\bibfnamefont {D.}~\bibnamefont
  {Becker}}, \bibinfo {author} {\bibfnamefont {H.}~\bibnamefont {Müntinga}},
  \bibinfo {author} {\bibfnamefont {H.}~\bibnamefont {Ahlers}}, \bibinfo
  {author} {\bibfnamefont {A.}~\bibnamefont {Peters}}, \bibinfo {author}
  {\bibfnamefont {C.}~\bibnamefont {Lämmerzahl}}, \bibinfo {author}
  {\bibfnamefont {K.}~\bibnamefont {Sengstock}}, \bibinfo {author}
  {\bibfnamefont {N.}~\bibnamefont {Gaaloul}}, \bibinfo {author} {\bibfnamefont
  {W.}~\bibnamefont {Ertmer}}, \ and\ \bibinfo {author} {\bibfnamefont {E.~M.}\
  \bibnamefont {Rasel}},\ }\href {\doibase 10.1088/1367-2630/17/6/065001}
  {\bibfield  {journal} {\bibinfo  {journal} {New Journal of Physics}\ }\textbf
  {\bibinfo {volume} {17}},\ \bibinfo {pages} {065001} (\bibinfo {year}
  {2015})}\BibitemShut {NoStop}%
\bibitem [{\citenamefont {Becker}\ \emph {et~al.}(2018)\citenamefont {Becker},
  \citenamefont {Lachmann}, \citenamefont {Seidel}, \citenamefont {Ahlers},
  \citenamefont {Dinkelaker}, \citenamefont {Grosse}, \citenamefont {Hellmig},
  \citenamefont {Müntinga}, \citenamefont {Schkolnik}, \citenamefont
  {Wendrich}, \citenamefont {Wenzlawski}, \citenamefont {Weps}, \citenamefont
  {Corgier}, \citenamefont {Franz}, \citenamefont {Gaaloul}, \citenamefont
  {Herr}, \citenamefont {Lüdtke}, \citenamefont {Popp}, \citenamefont {Amri},
  \citenamefont {Duncker}, \citenamefont {Erbe}, \citenamefont {Kohfeldt},
  \citenamefont {Kubelka-Lange}, \citenamefont {Braxmaier}, \citenamefont
  {Charron}, \citenamefont {Ertmer}, \citenamefont {Krutzik}, \citenamefont
  {Lämmerzahl}, \citenamefont {Peters}, \citenamefont {Schleich},
  \citenamefont {Sengstock}, \citenamefont {Walser}, \citenamefont {Wicht},
  \citenamefont {Windpassinger},\ and\ \citenamefont {Rasel}}]{Becker2018}%
  \BibitemOpen
  \bibfield  {author} {\bibinfo {author} {\bibfnamefont {D.}~\bibnamefont
  {Becker}}, \bibinfo {author} {\bibfnamefont {M.~D.}\ \bibnamefont
  {Lachmann}}, \bibinfo {author} {\bibfnamefont {S.~T.}\ \bibnamefont
  {Seidel}}, \bibinfo {author} {\bibfnamefont {H.}~\bibnamefont {Ahlers}},
  \bibinfo {author} {\bibfnamefont {A.~N.}\ \bibnamefont {Dinkelaker}},
  \bibinfo {author} {\bibfnamefont {J.}~\bibnamefont {Grosse}}, \bibinfo
  {author} {\bibfnamefont {O.}~\bibnamefont {Hellmig}}, \bibinfo {author}
  {\bibfnamefont {H.}~\bibnamefont {Müntinga}}, \bibinfo {author}
  {\bibfnamefont {V.}~\bibnamefont {Schkolnik}}, \bibinfo {author}
  {\bibfnamefont {T.}~\bibnamefont {Wendrich}}, \bibinfo {author}
  {\bibfnamefont {A.}~\bibnamefont {Wenzlawski}}, \bibinfo {author}
  {\bibfnamefont {B.}~\bibnamefont {Weps}}, \bibinfo {author} {\bibfnamefont
  {R.}~\bibnamefont {Corgier}}, \bibinfo {author} {\bibfnamefont
  {T.}~\bibnamefont {Franz}}, \bibinfo {author} {\bibfnamefont
  {N.}~\bibnamefont {Gaaloul}}, \bibinfo {author} {\bibfnamefont
  {W.}~\bibnamefont {Herr}}, \bibinfo {author} {\bibfnamefont {D.}~\bibnamefont
  {Lüdtke}}, \bibinfo {author} {\bibfnamefont {M.}~\bibnamefont {Popp}},
  \bibinfo {author} {\bibfnamefont {S.}~\bibnamefont {Amri}}, \bibinfo {author}
  {\bibfnamefont {H.}~\bibnamefont {Duncker}}, \bibinfo {author} {\bibfnamefont
  {M.}~\bibnamefont {Erbe}}, \bibinfo {author} {\bibfnamefont {A.}~\bibnamefont
  {Kohfeldt}}, \bibinfo {author} {\bibfnamefont {A.}~\bibnamefont
  {Kubelka-Lange}}, \bibinfo {author} {\bibfnamefont {C.}~\bibnamefont
  {Braxmaier}}, \bibinfo {author} {\bibfnamefont {E.}~\bibnamefont {Charron}},
  \bibinfo {author} {\bibfnamefont {W.}~\bibnamefont {Ertmer}}, \bibinfo
  {author} {\bibfnamefont {M.}~\bibnamefont {Krutzik}}, \bibinfo {author}
  {\bibfnamefont {C.}~\bibnamefont {Lämmerzahl}}, \bibinfo {author}
  {\bibfnamefont {A.}~\bibnamefont {Peters}}, \bibinfo {author} {\bibfnamefont
  {W.~P.}\ \bibnamefont {Schleich}}, \bibinfo {author} {\bibfnamefont
  {K.}~\bibnamefont {Sengstock}}, \bibinfo {author} {\bibfnamefont
  {R.}~\bibnamefont {Walser}}, \bibinfo {author} {\bibfnamefont
  {A.}~\bibnamefont {Wicht}}, \bibinfo {author} {\bibfnamefont
  {P.}~\bibnamefont {Windpassinger}}, \ and\ \bibinfo {author} {\bibfnamefont
  {E.~M.}\ \bibnamefont {Rasel}},\ }\href {\doibase 10.1038/s41586-018-0605-1}
  {\bibfield  {journal} {\bibinfo  {journal} {Nature}\ }\textbf {\bibinfo
  {volume} {562}},\ \bibinfo {pages} {391} (\bibinfo {year}
  {2018})}\BibitemShut {NoStop}%
\bibitem [{\citenamefont {O'Hara}\ \emph {et~al.}(2001)\citenamefont {O'Hara},
  \citenamefont {Gehm}, \citenamefont {Granade},\ and\ \citenamefont
  {Thomas}}]{OHara01PRA}%
  \BibitemOpen
  \bibfield  {author} {\bibinfo {author} {\bibfnamefont {K.~M.}\ \bibnamefont
  {O'Hara}}, \bibinfo {author} {\bibfnamefont {M.~E.}\ \bibnamefont {Gehm}},
  \bibinfo {author} {\bibfnamefont {S.~R.}\ \bibnamefont {Granade}}, \ and\
  \bibinfo {author} {\bibfnamefont {J.~E.}\ \bibnamefont {Thomas}},\ }\href
  {\doibase 10.1103/physreva.64.051403} {\bibfield  {journal} {\bibinfo
  {journal} {Physical Review A}\ }\textbf {\bibinfo {volume} {64}},\ \bibinfo
  {pages} {051403} (\bibinfo {year} {2001})}\BibitemShut {NoStop}%
\bibitem [{\citenamefont {Kinoshita}\ \emph {et~al.}(2005)\citenamefont
  {Kinoshita}, \citenamefont {Wenger},\ and\ \citenamefont
  {Weiss}}]{Kinoshita05PRAR}%
  \BibitemOpen
  \bibfield  {author} {\bibinfo {author} {\bibfnamefont {T.}~\bibnamefont
  {Kinoshita}}, \bibinfo {author} {\bibfnamefont {T.}~\bibnamefont {Wenger}}, \
  and\ \bibinfo {author} {\bibfnamefont {D.}~\bibnamefont {Weiss}},\ }\href
  {\doibase 10.1103/PhysRevA.71.011602} {\bibfield  {journal} {\bibinfo
  {journal} {Phys. Rev. A}\ }\textbf {\bibinfo {volume} {71}},\ \bibinfo
  {pages} {011602} (\bibinfo {year} {2005})}\BibitemShut {NoStop}%
\bibitem [{\citenamefont {Roy}\ \emph {et~al.}(2016)\citenamefont {Roy},
  \citenamefont {Green}, \citenamefont {Bowler},\ and\ \citenamefont
  {Gupta}}]{Roy2016PRA}%
  \BibitemOpen
  \bibfield  {author} {\bibinfo {author} {\bibfnamefont {R.}~\bibnamefont
  {Roy}}, \bibinfo {author} {\bibfnamefont {A.}~\bibnamefont {Green}}, \bibinfo
  {author} {\bibfnamefont {R.}~\bibnamefont {Bowler}}, \ and\ \bibinfo {author}
  {\bibfnamefont {S.}~\bibnamefont {Gupta}},\ }\href {\doibase
  10.1103/PhysRevA.93.043403} {\bibfield  {journal} {\bibinfo  {journal} {Phys.
  Rev. A}\ }\textbf {\bibinfo {volume} {93}},\ \bibinfo {pages} {043403}
  (\bibinfo {year} {2016})}\BibitemShut {NoStop}%
\bibitem [{\citenamefont {Stellmer}\ \emph {et~al.}(2013)\citenamefont
  {Stellmer}, \citenamefont {Grimm},\ and\ \citenamefont
  {Schreck}}]{Stellmer13PRA}%
  \BibitemOpen
  \bibfield  {author} {\bibinfo {author} {\bibfnamefont {S.}~\bibnamefont
  {Stellmer}}, \bibinfo {author} {\bibfnamefont {R.}~\bibnamefont {Grimm}}, \
  and\ \bibinfo {author} {\bibfnamefont {F.}~\bibnamefont {Schreck}},\ }\href
  {\doibase 10.1103/PhysRevA.87.013611} {\bibfield  {journal} {\bibinfo
  {journal} {Phys. Rev. A}\ }\textbf {\bibinfo {volume} {87}},\ \bibinfo
  {pages} {013611} (\bibinfo {year} {2013})}\BibitemShut {NoStop}%
\bibitem [{\citenamefont {Urvoy}\ \emph {et~al.}(2019)\citenamefont {Urvoy},
  \citenamefont {Vendeiro}, \citenamefont {Ramette}, \citenamefont
  {Adiyatullin},\ and\ \citenamefont {Vuleti\ifmmode~\acute{c}\else
  \'{c}\fi{}}}]{Urvoy19PRL}%
  \BibitemOpen
  \bibfield  {author} {\bibinfo {author} {\bibfnamefont {A.}~\bibnamefont
  {Urvoy}}, \bibinfo {author} {\bibfnamefont {Z.}~\bibnamefont {Vendeiro}},
  \bibinfo {author} {\bibfnamefont {J.}~\bibnamefont {Ramette}}, \bibinfo
  {author} {\bibfnamefont {A.}~\bibnamefont {Adiyatullin}}, \ and\ \bibinfo
  {author} {\bibfnamefont {V.}~\bibnamefont {Vuleti\ifmmode~\acute{c}\else
  \'{c}\fi{}}},\ }\href {\doibase 10.1103/PhysRevLett.122.203202} {\bibfield
  {journal} {\bibinfo  {journal} {Phys. Rev. Lett.}\ }\textbf {\bibinfo
  {volume} {122}},\ \bibinfo {pages} {203202} (\bibinfo {year}
  {2019})}\BibitemShut {NoStop}%
\bibitem [{\citenamefont {Vendeiro}\ \emph {et~al.}(2022)\citenamefont
  {Vendeiro}, \citenamefont {Ramette}, \citenamefont {Rudelis}, \citenamefont
  {Chong}, \citenamefont {Sinclair}, \citenamefont {Stewart}, \citenamefont
  {Urvoy},\ and\ \citenamefont {Vuleti\ifmmode~\acute{c}\else
  \'{c}\fi{}}}]{Vendeiro2022PRR}%
  \BibitemOpen
  \bibfield  {author} {\bibinfo {author} {\bibfnamefont {Z.}~\bibnamefont
  {Vendeiro}}, \bibinfo {author} {\bibfnamefont {J.}~\bibnamefont {Ramette}},
  \bibinfo {author} {\bibfnamefont {A.}~\bibnamefont {Rudelis}}, \bibinfo
  {author} {\bibfnamefont {M.}~\bibnamefont {Chong}}, \bibinfo {author}
  {\bibfnamefont {J.}~\bibnamefont {Sinclair}}, \bibinfo {author}
  {\bibfnamefont {L.}~\bibnamefont {Stewart}}, \bibinfo {author} {\bibfnamefont
  {A.}~\bibnamefont {Urvoy}}, \ and\ \bibinfo {author} {\bibfnamefont
  {V.}~\bibnamefont {Vuleti\ifmmode~\acute{c}\else \'{c}\fi{}}},\ }\href
  {\doibase 10.1103/PhysRevResearch.4.043216} {\bibfield  {journal} {\bibinfo
  {journal} {Phys. Rev. Res.}\ }\textbf {\bibinfo {volume} {4}},\ \bibinfo
  {pages} {043216} (\bibinfo {year} {2022})}\BibitemShut {NoStop}%
\bibitem [{\citenamefont {Inouye}\ \emph {et~al.}(1998)\citenamefont {Inouye},
  \citenamefont {Andrews}, \citenamefont {Stenger}, \citenamefont {Miesner},
  \citenamefont {Stamper-Kurn},\ and\ \citenamefont
  {Ketterle}}]{Inouye98Nature}%
  \BibitemOpen
  \bibfield  {author} {\bibinfo {author} {\bibfnamefont {S.}~\bibnamefont
  {Inouye}}, \bibinfo {author} {\bibfnamefont {M.~R.}\ \bibnamefont {Andrews}},
  \bibinfo {author} {\bibfnamefont {J.}~\bibnamefont {Stenger}}, \bibinfo
  {author} {\bibfnamefont {H.-J.}\ \bibnamefont {Miesner}}, \bibinfo {author}
  {\bibfnamefont {D.~M.}\ \bibnamefont {Stamper-Kurn}}, \ and\ \bibinfo
  {author} {\bibfnamefont {W.}~\bibnamefont {Ketterle}},\ }\href {\doibase
  10.1038/32354} {\bibfield  {journal} {\bibinfo  {journal} {Nature}\ }\textbf
  {\bibinfo {volume} {392}},\ \bibinfo {pages} {151} (\bibinfo {year}
  {1998})}\BibitemShut {NoStop}%
\bibitem [{\citenamefont {D{\textquotesingle}Errico}\ \emph
  {et~al.}(2007)\citenamefont {D{\textquotesingle}Errico}, \citenamefont
  {Zaccanti}, \citenamefont {Fattori}, \citenamefont {Roati}, \citenamefont
  {Inguscio}, \citenamefont {Modugno},\ and\ \citenamefont
  {Simoni}}]{DErrico07NJP}%
  \BibitemOpen
  \bibfield  {author} {\bibinfo {author} {\bibfnamefont {C.}~\bibnamefont
  {D{\textquotesingle}Errico}}, \bibinfo {author} {\bibfnamefont
  {M.}~\bibnamefont {Zaccanti}}, \bibinfo {author} {\bibfnamefont
  {M.}~\bibnamefont {Fattori}}, \bibinfo {author} {\bibfnamefont
  {G.}~\bibnamefont {Roati}}, \bibinfo {author} {\bibfnamefont
  {M.}~\bibnamefont {Inguscio}}, \bibinfo {author} {\bibfnamefont
  {G.}~\bibnamefont {Modugno}}, \ and\ \bibinfo {author} {\bibfnamefont
  {A.}~\bibnamefont {Simoni}},\ }\href {\doibase 10.1088/1367-2630/9/7/223}
  {\bibfield  {journal} {\bibinfo  {journal} {New Journal of Physics}\ }\textbf
  {\bibinfo {volume} {9}},\ \bibinfo {pages} {223} (\bibinfo {year}
  {2007})}\BibitemShut {NoStop}%
\bibitem [{\citenamefont {Salomon}\ \emph {et~al.}(2013)\citenamefont
  {Salomon}, \citenamefont {Fouch{\'e}}, \citenamefont {Wang}, \citenamefont
  {Aspect}, \citenamefont {Bouyer},\ and\ \citenamefont
  {Bourdel}}]{Salomon13EPL}%
  \BibitemOpen
  \bibfield  {author} {\bibinfo {author} {\bibfnamefont {G.}~\bibnamefont
  {Salomon}}, \bibinfo {author} {\bibfnamefont {L.}~\bibnamefont {Fouch{\'e}}},
  \bibinfo {author} {\bibfnamefont {P.}~\bibnamefont {Wang}}, \bibinfo {author}
  {\bibfnamefont {A.}~\bibnamefont {Aspect}}, \bibinfo {author} {\bibfnamefont
  {P.}~\bibnamefont {Bouyer}}, \ and\ \bibinfo {author} {\bibfnamefont
  {T.}~\bibnamefont {Bourdel}},\ }\href
  {http://stacks.iop.org/0295-5075/104/i=6/a=63002} {\bibfield  {journal}
  {\bibinfo  {journal} {Eur. Phys. Lett.}\ }\textbf {\bibinfo {volume} {104}},\
  \bibinfo {pages} {63002} (\bibinfo {year} {2013})}\BibitemShut {NoStop}%
\bibitem [{\citenamefont {Herbst}\ \emph {et~al.}(2022)\citenamefont {Herbst},
  \citenamefont {Albers}, \citenamefont {Stolzenberg}, \citenamefont {Bode},\
  and\ \citenamefont {Schlippert}}]{Herbst2022PRA}%
  \BibitemOpen
  \bibfield  {author} {\bibinfo {author} {\bibfnamefont {A.}~\bibnamefont
  {Herbst}}, \bibinfo {author} {\bibfnamefont {H.}~\bibnamefont {Albers}},
  \bibinfo {author} {\bibfnamefont {K.}~\bibnamefont {Stolzenberg}}, \bibinfo
  {author} {\bibfnamefont {S.}~\bibnamefont {Bode}}, \ and\ \bibinfo {author}
  {\bibfnamefont {D.}~\bibnamefont {Schlippert}},\ }\href {\doibase
  10.1103/PhysRevA.106.043320} {\bibfield  {journal} {\bibinfo  {journal}
  {Phys. Rev. A}\ }\textbf {\bibinfo {volume} {106}},\ \bibinfo {pages}
  {043320} (\bibinfo {year} {2022})}\BibitemShut {NoStop}%
\bibitem [{\citenamefont {Landini}\ \emph {et~al.}(2012)\citenamefont
  {Landini}, \citenamefont {Roy}, \citenamefont {Roati}, \citenamefont
  {Simoni}, \citenamefont {Inguscio}, \citenamefont {Modugno},\ and\
  \citenamefont {Fattori}}]{Landini12PRA}%
  \BibitemOpen
  \bibfield  {author} {\bibinfo {author} {\bibfnamefont {M.}~\bibnamefont
  {Landini}}, \bibinfo {author} {\bibfnamefont {S.}~\bibnamefont {Roy}},
  \bibinfo {author} {\bibfnamefont {G.}~\bibnamefont {Roati}}, \bibinfo
  {author} {\bibfnamefont {A.}~\bibnamefont {Simoni}}, \bibinfo {author}
  {\bibfnamefont {M.}~\bibnamefont {Inguscio}}, \bibinfo {author}
  {\bibfnamefont {G.}~\bibnamefont {Modugno}}, \ and\ \bibinfo {author}
  {\bibfnamefont {M.}~\bibnamefont {Fattori}},\ }\href {\doibase
  10.1103/PhysRevA.86.033421} {\bibfield  {journal} {\bibinfo  {journal} {Phys.
  Rev. A}\ }\textbf {\bibinfo {volume} {86}},\ \bibinfo {pages} {033421}
  (\bibinfo {year} {2012})}\BibitemShut {NoStop}%
\bibitem [{\citenamefont {Tiemann}\ \emph {et~al.}(2020)\citenamefont
  {Tiemann}, \citenamefont {Gersema}, \citenamefont {Voges}, \citenamefont
  {Hartmann}, \citenamefont {Zenesini},\ and\ \citenamefont
  {Ospelkaus}}]{Tiemann2020PRR}%
  \BibitemOpen
  \bibfield  {author} {\bibinfo {author} {\bibfnamefont {E.}~\bibnamefont
  {Tiemann}}, \bibinfo {author} {\bibfnamefont {P.}~\bibnamefont {Gersema}},
  \bibinfo {author} {\bibfnamefont {K.~K.}\ \bibnamefont {Voges}}, \bibinfo
  {author} {\bibfnamefont {T.}~\bibnamefont {Hartmann}}, \bibinfo {author}
  {\bibfnamefont {A.}~\bibnamefont {Zenesini}}, \ and\ \bibinfo {author}
  {\bibfnamefont {S.}~\bibnamefont {Ospelkaus}},\ }\href {\doibase
  10.1103/PhysRevResearch.2.013366} {\bibfield  {journal} {\bibinfo  {journal}
  {Phys. Rev. Res.}\ }\textbf {\bibinfo {volume} {2}},\ \bibinfo {pages}
  {013366} (\bibinfo {year} {2020})}\BibitemShut {NoStop}%
\bibitem [{\citenamefont {Ketterle}\ and\ \citenamefont
  {Druten}(1996)}]{Ketterle96AAMOP}%
  \BibitemOpen
  \bibfield  {author} {\bibinfo {author} {\bibfnamefont {W.}~\bibnamefont
  {Ketterle}}\ and\ \bibinfo {author} {\bibfnamefont {N.~V.}\ \bibnamefont
  {Druten}},\ }in\ \href {\doibase 10.1016/s1049-250x(08)60101-9} {\emph
  {\bibinfo {booktitle} {Advances In Atomic, Molecular, and Optical Physics}}}\
  (\bibinfo  {publisher} {Elsevier},\ \bibinfo {year} {1996})\ pp.\ \bibinfo
  {pages} {181--236}\BibitemShut {NoStop}%
\bibitem [{\citenamefont {Monroe}\ \emph {et~al.}(1993)\citenamefont {Monroe},
  \citenamefont {Cornell}, \citenamefont {Sackett}, \citenamefont {Myatt},\
  and\ \citenamefont {Wieman}}]{Monroe1993PRL}%
  \BibitemOpen
  \bibfield  {author} {\bibinfo {author} {\bibfnamefont {C.~R.}\ \bibnamefont
  {Monroe}}, \bibinfo {author} {\bibfnamefont {E.~A.}\ \bibnamefont {Cornell}},
  \bibinfo {author} {\bibfnamefont {C.~A.}\ \bibnamefont {Sackett}}, \bibinfo
  {author} {\bibfnamefont {C.~J.}\ \bibnamefont {Myatt}}, \ and\ \bibinfo
  {author} {\bibfnamefont {C.~E.}\ \bibnamefont {Wieman}},\ }\href {\doibase
  10.1103/PhysRevLett.70.414} {\bibfield  {journal} {\bibinfo  {journal} {Phys.
  Rev. Lett.}\ }\textbf {\bibinfo {volume} {70}},\ \bibinfo {pages} {414}
  (\bibinfo {year} {1993})}\BibitemShut {NoStop}%
\bibitem [{\citenamefont {Weber}\ \emph
  {et~al.}(2003{\natexlab{a}})\citenamefont {Weber}, \citenamefont {Herbig},
  \citenamefont {Mark}, \citenamefont {N\"agerl},\ and\ \citenamefont
  {Grimm}}]{Weber03PRL}%
  \BibitemOpen
  \bibfield  {author} {\bibinfo {author} {\bibfnamefont {T.}~\bibnamefont
  {Weber}}, \bibinfo {author} {\bibfnamefont {J.}~\bibnamefont {Herbig}},
  \bibinfo {author} {\bibfnamefont {M.}~\bibnamefont {Mark}}, \bibinfo {author}
  {\bibfnamefont {H.-C.}\ \bibnamefont {N\"agerl}}, \ and\ \bibinfo {author}
  {\bibfnamefont {R.}~\bibnamefont {Grimm}},\ }\href {\doibase
  10.1103/PhysRevLett.91.123201} {\bibfield  {journal} {\bibinfo  {journal}
  {Phys. Rev. Lett.}\ }\textbf {\bibinfo {volume} {91}},\ \bibinfo {pages}
  {123201} (\bibinfo {year} {2003}{\natexlab{a}})}\BibitemShut {NoStop}%
\bibitem [{\citenamefont {Weber}\ \emph
  {et~al.}(2003{\natexlab{b}})\citenamefont {Weber}, \citenamefont {Herbig},
  \citenamefont {Mark}, \citenamefont {N{\"a}gerl},\ and\ \citenamefont
  {Grimm}}]{Weber2003Science}%
  \BibitemOpen
  \bibfield  {author} {\bibinfo {author} {\bibfnamefont {T.}~\bibnamefont
  {Weber}}, \bibinfo {author} {\bibfnamefont {J.}~\bibnamefont {Herbig}},
  \bibinfo {author} {\bibfnamefont {M.}~\bibnamefont {Mark}}, \bibinfo {author}
  {\bibfnamefont {H.-C.}\ \bibnamefont {N{\"a}gerl}}, \ and\ \bibinfo {author}
  {\bibfnamefont {R.}~\bibnamefont {Grimm}},\ }\href {\doibase
  10.1126/science.1079699} {\bibfield  {journal} {\bibinfo  {journal} {Science
  (New York, N.Y.)}\ }\textbf {\bibinfo {volume} {299}},\ \bibinfo {pages}
  {232} (\bibinfo {year} {2003}{\natexlab{b}})}\BibitemShut {NoStop}%
\bibitem [{\citenamefont {Kraemer}\ \emph {et~al.}(2004)\citenamefont
  {Kraemer}, \citenamefont {Herbig}, \citenamefont {Mark}, \citenamefont
  {Weber}, \citenamefont {Chin}, \citenamefont {N{\"a}gerl},\ and\
  \citenamefont {Grimm}}]{Kraemer2004APB}%
  \BibitemOpen
  \bibfield  {author} {\bibinfo {author} {\bibfnamefont {T.}~\bibnamefont
  {Kraemer}}, \bibinfo {author} {\bibfnamefont {J.}~\bibnamefont {Herbig}},
  \bibinfo {author} {\bibfnamefont {M.}~\bibnamefont {Mark}}, \bibinfo {author}
  {\bibfnamefont {T.}~\bibnamefont {Weber}}, \bibinfo {author} {\bibfnamefont
  {C.}~\bibnamefont {Chin}}, \bibinfo {author} {\bibfnamefont {H.-C.}\
  \bibnamefont {N{\"a}gerl}}, \ and\ \bibinfo {author} {\bibfnamefont
  {R.}~\bibnamefont {Grimm}},\ }\href {\doibase 10.1007/s00340-004-1657-5}
  {\bibfield  {journal} {\bibinfo  {journal} {Applied Physics B Photophysics
  and Laser Chemistry}\ }\textbf {\bibinfo {volume} {79}},\ \bibinfo {pages}
  {1013} (\bibinfo {year} {2004})}\BibitemShut {NoStop}%
\bibitem [{\citenamefont {Roati}\ \emph {et~al.}(2007)\citenamefont {Roati},
  \citenamefont {Zaccanti}, \citenamefont {D'Errico}, \citenamefont {Catani},
  \citenamefont {Modugno}, \citenamefont {Simoni}, \citenamefont {Inguscio},\
  and\ \citenamefont {Modugno}}]{Roati2007PRL}%
  \BibitemOpen
  \bibfield  {author} {\bibinfo {author} {\bibfnamefont {G.}~\bibnamefont
  {Roati}}, \bibinfo {author} {\bibfnamefont {M.}~\bibnamefont {Zaccanti}},
  \bibinfo {author} {\bibfnamefont {C.}~\bibnamefont {D'Errico}}, \bibinfo
  {author} {\bibfnamefont {J.}~\bibnamefont {Catani}}, \bibinfo {author}
  {\bibfnamefont {M.}~\bibnamefont {Modugno}}, \bibinfo {author} {\bibfnamefont
  {A.}~\bibnamefont {Simoni}}, \bibinfo {author} {\bibfnamefont
  {M.}~\bibnamefont {Inguscio}}, \ and\ \bibinfo {author} {\bibfnamefont
  {G.}~\bibnamefont {Modugno}},\ }\href {\doibase
  10.1103/PhysRevLett.99.010403} {\bibfield  {journal} {\bibinfo  {journal}
  {Phys. Rev. Lett.}\ }\textbf {\bibinfo {volume} {99}},\ \bibinfo {pages}
  {010403} (\bibinfo {year} {2007})}\BibitemShut {NoStop}%
\bibitem [{\citenamefont {Castin}\ and\ \citenamefont
  {Dum}(1996)}]{Castin96PRL}%
  \BibitemOpen
  \bibfield  {author} {\bibinfo {author} {\bibfnamefont {Y.}~\bibnamefont
  {Castin}}\ and\ \bibinfo {author} {\bibfnamefont {R.}~\bibnamefont {Dum}},\
  }\href {\doibase 10.1103/PhysRevLett.77.5315} {\bibfield  {journal} {\bibinfo
   {journal} {Phys. Rev. Lett.}\ }\textbf {\bibinfo {volume} {77}},\ \bibinfo
  {pages} {5315} (\bibinfo {year} {1996})}\BibitemShut {NoStop}%
\bibitem [{\citenamefont {Kagan}\ \emph {et~al.}(1997)\citenamefont {Kagan},
  \citenamefont {Surkov},\ and\ \citenamefont {Shlyapnikov}}]{Kagan97PRA}%
  \BibitemOpen
  \bibfield  {author} {\bibinfo {author} {\bibfnamefont {Y.}~\bibnamefont
  {Kagan}}, \bibinfo {author} {\bibfnamefont {E.~L.}\ \bibnamefont {Surkov}}, \
  and\ \bibinfo {author} {\bibfnamefont {G.~V.}\ \bibnamefont {Shlyapnikov}},\
  }\href {\doibase 10.1103/PhysRevA.55.R18} {\bibfield  {journal} {\bibinfo
  {journal} {Phys. Rev. A}\ }\textbf {\bibinfo {volume} {55}},\ \bibinfo
  {pages} {R18} (\bibinfo {year} {1997})}\BibitemShut {NoStop}%
\bibitem [{\citenamefont {Corgier}\ \emph {et~al.}(2020)\citenamefont
  {Corgier}, \citenamefont {Loriani}, \citenamefont {Ahlers}, \citenamefont
  {Posso-Trujillo}, \citenamefont {Schubert}, \citenamefont {Rasel},
  \citenamefont {Charron},\ and\ \citenamefont {Gaaloul}}]{Corgier_2020}%
  \BibitemOpen
  \bibfield  {author} {\bibinfo {author} {\bibfnamefont {R.}~\bibnamefont
  {Corgier}}, \bibinfo {author} {\bibfnamefont {S.}~\bibnamefont {Loriani}},
  \bibinfo {author} {\bibfnamefont {H.}~\bibnamefont {Ahlers}}, \bibinfo
  {author} {\bibfnamefont {K.}~\bibnamefont {Posso-Trujillo}}, \bibinfo
  {author} {\bibfnamefont {C.}~\bibnamefont {Schubert}}, \bibinfo {author}
  {\bibfnamefont {E.~M.}\ \bibnamefont {Rasel}}, \bibinfo {author}
  {\bibfnamefont {E.}~\bibnamefont {Charron}}, \ and\ \bibinfo {author}
  {\bibfnamefont {N.}~\bibnamefont {Gaaloul}},\ }\href {\doibase
  10.1088/1367-2630/abcbc8} {\bibfield  {journal} {\bibinfo  {journal} {New
  Journal of Physics}\ }\textbf {\bibinfo {volume} {22}},\ \bibinfo {pages}
  {123008} (\bibinfo {year} {2020})}\BibitemShut {NoStop}%
\bibitem [{\citenamefont {P\'erez-Garc\'{\i}a}\ \emph
  {et~al.}(1996)\citenamefont {P\'erez-Garc\'{\i}a}, \citenamefont {Michinel},
  \citenamefont {Cirac}, \citenamefont {Lewenstein},\ and\ \citenamefont
  {Zoller}}]{Perez96PRL}%
  \BibitemOpen
  \bibfield  {author} {\bibinfo {author} {\bibfnamefont {V.~M.}\ \bibnamefont
  {P\'erez-Garc\'{\i}a}}, \bibinfo {author} {\bibfnamefont {H.}~\bibnamefont
  {Michinel}}, \bibinfo {author} {\bibfnamefont {J.~I.}\ \bibnamefont {Cirac}},
  \bibinfo {author} {\bibfnamefont {M.}~\bibnamefont {Lewenstein}}, \ and\
  \bibinfo {author} {\bibfnamefont {P.}~\bibnamefont {Zoller}},\ }\href
  {\doibase 10.1103/PhysRevLett.77.5320} {\bibfield  {journal} {\bibinfo
  {journal} {Phys. Rev. Lett.}\ }\textbf {\bibinfo {volume} {77}},\ \bibinfo
  {pages} {5320} (\bibinfo {year} {1996})}\BibitemShut {NoStop}%
\bibitem [{\citenamefont {P\'erez-Garc\'{\i}a}\ \emph
  {et~al.}(1997)\citenamefont {P\'erez-Garc\'{\i}a}, \citenamefont {Michinel},
  \citenamefont {Cirac}, \citenamefont {Lewenstein},\ and\ \citenamefont
  {Zoller}}]{Perez97PRA}%
  \BibitemOpen
  \bibfield  {author} {\bibinfo {author} {\bibfnamefont {V.~M.}\ \bibnamefont
  {P\'erez-Garc\'{\i}a}}, \bibinfo {author} {\bibfnamefont {H.}~\bibnamefont
  {Michinel}}, \bibinfo {author} {\bibfnamefont {J.~I.}\ \bibnamefont {Cirac}},
  \bibinfo {author} {\bibfnamefont {M.}~\bibnamefont {Lewenstein}}, \ and\
  \bibinfo {author} {\bibfnamefont {P.}~\bibnamefont {Zoller}},\ }\href
  {\doibase 10.1103/PhysRevA.56.1424} {\bibfield  {journal} {\bibinfo
  {journal} {Phys. Rev. A}\ }\textbf {\bibinfo {volume} {56}},\ \bibinfo
  {pages} {1424} (\bibinfo {year} {1997})}\BibitemShut {NoStop}%
\bibitem [{\citenamefont {Albers}\ \emph {et~al.}(2022)\citenamefont {Albers},
  \citenamefont {Corgier}, \citenamefont {Herbst}, \citenamefont {Rajagopalan},
  \citenamefont {Schubert}, \citenamefont {Vogt}, \citenamefont {Woltmann},
  \citenamefont {L{\"a}mmerzahl}, \citenamefont {Herrmann}, \citenamefont
  {Charron}, \citenamefont {Ertmer}, \citenamefont {Rasel}, \citenamefont
  {Gaaloul},\ and\ \citenamefont {Schlippert}}]{Albers2022Commun}%
  \BibitemOpen
  \bibfield  {author} {\bibinfo {author} {\bibfnamefont {H.}~\bibnamefont
  {Albers}}, \bibinfo {author} {\bibfnamefont {R.}~\bibnamefont {Corgier}},
  \bibinfo {author} {\bibfnamefont {A.}~\bibnamefont {Herbst}}, \bibinfo
  {author} {\bibfnamefont {A.}~\bibnamefont {Rajagopalan}}, \bibinfo {author}
  {\bibfnamefont {C.}~\bibnamefont {Schubert}}, \bibinfo {author}
  {\bibfnamefont {C.}~\bibnamefont {Vogt}}, \bibinfo {author} {\bibfnamefont
  {M.}~\bibnamefont {Woltmann}}, \bibinfo {author} {\bibfnamefont
  {C.}~\bibnamefont {L{\"a}mmerzahl}}, \bibinfo {author} {\bibfnamefont
  {S.}~\bibnamefont {Herrmann}}, \bibinfo {author} {\bibfnamefont
  {E.}~\bibnamefont {Charron}}, \bibinfo {author} {\bibfnamefont
  {W.}~\bibnamefont {Ertmer}}, \bibinfo {author} {\bibfnamefont {E.~M.}\
  \bibnamefont {Rasel}}, \bibinfo {author} {\bibfnamefont {N.}~\bibnamefont
  {Gaaloul}}, \ and\ \bibinfo {author} {\bibfnamefont {D.}~\bibnamefont
  {Schlippert}},\ }\href {\doibase 10.1038/s42005-022-00825-2} {\bibfield
  {journal} {\bibinfo  {journal} {Communications Physics}\ }\textbf {\bibinfo
  {volume} {5}} (\bibinfo {year} {2022}),\
  10.1038/s42005-022-00825-2}\BibitemShut {NoStop}%
\bibitem [{\citenamefont {Loriani}\ \emph {et~al.}()\citenamefont {Loriani},
  \citenamefont {Schlippert}, \citenamefont {Schubert}, \citenamefont {Abend},
  \citenamefont {Ahlers}, \citenamefont {Ertmer}, \citenamefont {Rudolph},
  \citenamefont {Hogan}, \citenamefont {Kasevich}, \citenamefont {Rasel},\ and\
  \citenamefont {Gaaloul}}]{Loriani2019NJP}%
  \BibitemOpen
  \bibfield  {author} {\bibinfo {author} {\bibfnamefont {S.}~\bibnamefont
  {Loriani}}, \bibinfo {author} {\bibfnamefont {D.}~\bibnamefont {Schlippert}},
  \bibinfo {author} {\bibfnamefont {C.}~\bibnamefont {Schubert}}, \bibinfo
  {author} {\bibfnamefont {S.}~\bibnamefont {Abend}}, \bibinfo {author}
  {\bibfnamefont {H.}~\bibnamefont {Ahlers}}, \bibinfo {author} {\bibfnamefont
  {W.}~\bibnamefont {Ertmer}}, \bibinfo {author} {\bibfnamefont
  {J.}~\bibnamefont {Rudolph}}, \bibinfo {author} {\bibfnamefont {J.~M.}\
  \bibnamefont {Hogan}}, \bibinfo {author} {\bibfnamefont {M.~A.}\ \bibnamefont
  {Kasevich}}, \bibinfo {author} {\bibfnamefont {E.~M.}\ \bibnamefont {Rasel}},
  \ and\ \bibinfo {author} {\bibfnamefont {N.}~\bibnamefont {Gaaloul}},\ }\href
  {\doibase 10.1088/1367-2630/ab22d0} {\bibfield  {journal} {\bibinfo
  {journal} {New Journal of Physics}\ }\textbf {\bibinfo {volume} {21}},\
  \bibinfo {pages} {063030}}\BibitemShut {NoStop}%
\bibitem [{\citenamefont {Catani}\ \emph {et~al.}(2006)\citenamefont {Catani},
  \citenamefont {Maioli}, \citenamefont {De~Sarlo}, \citenamefont {Minardi},\
  and\ \citenamefont {Inguscio}}]{Catani2006PRA}%
  \BibitemOpen
  \bibfield  {author} {\bibinfo {author} {\bibfnamefont {J.}~\bibnamefont
  {Catani}}, \bibinfo {author} {\bibfnamefont {P.}~\bibnamefont {Maioli}},
  \bibinfo {author} {\bibfnamefont {L.}~\bibnamefont {De~Sarlo}}, \bibinfo
  {author} {\bibfnamefont {F.}~\bibnamefont {Minardi}}, \ and\ \bibinfo
  {author} {\bibfnamefont {M.}~\bibnamefont {Inguscio}},\ }\href {\doibase
  10.1103/PhysRevA.73.033415} {\bibfield  {journal} {\bibinfo  {journal} {Phys.
  Rev. A}\ }\textbf {\bibinfo {volume} {73}},\ \bibinfo {pages} {033415}
  (\bibinfo {year} {2006})}\BibitemShut {NoStop}%
\bibitem [{\citenamefont {Chaudhuri}\ \emph {et~al.}(2006)\citenamefont
  {Chaudhuri}, \citenamefont {Roy},\ and\ \citenamefont
  {Unnikrishnan}}]{Chaudhuri06PRA}%
  \BibitemOpen
  \bibfield  {author} {\bibinfo {author} {\bibfnamefont {S.}~\bibnamefont
  {Chaudhuri}}, \bibinfo {author} {\bibfnamefont {S.}~\bibnamefont {Roy}}, \
  and\ \bibinfo {author} {\bibfnamefont {C.~S.}\ \bibnamefont {Unnikrishnan}},\
  }\href {\doibase 10.1103/PhysRevA.74.023406} {\bibfield  {journal} {\bibinfo
  {journal} {Phys. Rev. A}\ }\textbf {\bibinfo {volume} {74}},\ \bibinfo
  {pages} {023406} (\bibinfo {year} {2006})}\BibitemShut {NoStop}%
\bibitem [{\citenamefont {Lasner}\ \emph {et~al.}(2021)\citenamefont {Lasner},
  \citenamefont {Mitra}, \citenamefont {Hiradfar}, \citenamefont {Augenbraun},
  \citenamefont {Cheuk}, \citenamefont {Lee}, \citenamefont {Prabhu},\ and\
  \citenamefont {Doyle}}]{Lasner21PRA}%
  \BibitemOpen
  \bibfield  {author} {\bibinfo {author} {\bibfnamefont {Z.}~\bibnamefont
  {Lasner}}, \bibinfo {author} {\bibfnamefont {D.}~\bibnamefont {Mitra}},
  \bibinfo {author} {\bibfnamefont {M.}~\bibnamefont {Hiradfar}}, \bibinfo
  {author} {\bibfnamefont {B.}~\bibnamefont {Augenbraun}}, \bibinfo {author}
  {\bibfnamefont {L.}~\bibnamefont {Cheuk}}, \bibinfo {author} {\bibfnamefont
  {E.}~\bibnamefont {Lee}}, \bibinfo {author} {\bibfnamefont {S.}~\bibnamefont
  {Prabhu}}, \ and\ \bibinfo {author} {\bibfnamefont {J.}~\bibnamefont
  {Doyle}},\ }\href {\doibase 10.1103/PhysRevA.104.063305} {\bibfield
  {journal} {\bibinfo  {journal} {Phys. Rev. A}\ }\textbf {\bibinfo {volume}
  {104}},\ \bibinfo {pages} {063305} (\bibinfo {year} {2021})}\BibitemShut
  {NoStop}%
\bibitem [{\citenamefont {Kovachy}\ \emph
  {et~al.}(2015{\natexlab{b}})\citenamefont {Kovachy}, \citenamefont {Hogan},
  \citenamefont {Sugarbaker}, \citenamefont {Dickerson}, \citenamefont
  {Donnelly}, \citenamefont {Overstreet},\ and\ \citenamefont
  {Kasevich}}]{Kovachy15PRL}%
  \BibitemOpen
  \bibfield  {author} {\bibinfo {author} {\bibfnamefont {T.}~\bibnamefont
  {Kovachy}}, \bibinfo {author} {\bibfnamefont {J.~M.}\ \bibnamefont {Hogan}},
  \bibinfo {author} {\bibfnamefont {A.}~\bibnamefont {Sugarbaker}}, \bibinfo
  {author} {\bibfnamefont {S.~M.}\ \bibnamefont {Dickerson}}, \bibinfo {author}
  {\bibfnamefont {C.~A.}\ \bibnamefont {Donnelly}}, \bibinfo {author}
  {\bibfnamefont {C.}~\bibnamefont {Overstreet}}, \ and\ \bibinfo {author}
  {\bibfnamefont {M.~A.}\ \bibnamefont {Kasevich}},\ }\href {\doibase
  10.1103/PhysRevLett.114.143004} {\bibfield  {journal} {\bibinfo  {journal}
  {Physical review letters}\ }\textbf {\bibinfo {volume} {114}},\ \bibinfo
  {pages} {143004} (\bibinfo {year} {2015}{\natexlab{b}})}\BibitemShut
  {NoStop}%
\bibitem [{\citenamefont {Hartwig}\ \emph {et~al.}(2015)\citenamefont
  {Hartwig}, \citenamefont {Abend}, \citenamefont {Schubert}, \citenamefont
  {Schlippert}, \citenamefont {Ahlers}, \citenamefont {Posso-Trujillo},
  \citenamefont {Gaaloul}, \citenamefont {Ertmer},\ and\ \citenamefont
  {Rasel}}]{Hartwig15NJP}%
  \BibitemOpen
  \bibfield  {author} {\bibinfo {author} {\bibfnamefont {J.}~\bibnamefont
  {Hartwig}}, \bibinfo {author} {\bibfnamefont {S.}~\bibnamefont {Abend}},
  \bibinfo {author} {\bibfnamefont {C.}~\bibnamefont {Schubert}}, \bibinfo
  {author} {\bibfnamefont {D.}~\bibnamefont {Schlippert}}, \bibinfo {author}
  {\bibfnamefont {H.}~\bibnamefont {Ahlers}}, \bibinfo {author} {\bibfnamefont
  {K.}~\bibnamefont {Posso-Trujillo}}, \bibinfo {author} {\bibfnamefont
  {N.}~\bibnamefont {Gaaloul}}, \bibinfo {author} {\bibfnamefont
  {W.}~\bibnamefont {Ertmer}}, \ and\ \bibinfo {author} {\bibfnamefont {E.~M.}\
  \bibnamefont {Rasel}},\ }\href {\doibase 10.1088/1367-2630/17/3/035011}
  {\bibfield  {journal} {\bibinfo  {journal} {New Journal of Physics}\ }\textbf
  {\bibinfo {volume} {17}},\ \bibinfo {pages} {035011} (\bibinfo {year}
  {2015})}\BibitemShut {NoStop}%
\bibitem [{\citenamefont {Schlippert}\ \emph {et~al.}(2020)\citenamefont
  {Schlippert}, \citenamefont {Meiners}, \citenamefont {Rengelink},
  \citenamefont {Schubert}, \citenamefont {Tell}, \citenamefont {Wodey},
  \citenamefont {Zipfel}, \citenamefont {Ertmer},\ and\ \citenamefont
  {Rasel}}]{Schlippert20WS}%
  \BibitemOpen
  \bibfield  {author} {\bibinfo {author} {\bibfnamefont {D.}~\bibnamefont
  {Schlippert}}, \bibinfo {author} {\bibfnamefont {C.}~\bibnamefont {Meiners}},
  \bibinfo {author} {\bibfnamefont {R.~J.}\ \bibnamefont {Rengelink}}, \bibinfo
  {author} {\bibfnamefont {C.}~\bibnamefont {Schubert}}, \bibinfo {author}
  {\bibfnamefont {D.}~\bibnamefont {Tell}}, \bibinfo {author} {\bibfnamefont
  {{\'E}.}~\bibnamefont {Wodey}}, \bibinfo {author} {\bibfnamefont {K.~H.}\
  \bibnamefont {Zipfel}}, \bibinfo {author} {\bibfnamefont {W.}~\bibnamefont
  {Ertmer}}, \ and\ \bibinfo {author} {\bibfnamefont {E.~M.}\ \bibnamefont
  {Rasel}},\ }in\ \href {\doibase 10.1142/9789811213984{\textunderscore }0010}
  {\emph {\bibinfo {booktitle} {CPT and Lorentz Symmetry}}},\ \bibinfo {editor}
  {edited by\ \bibinfo {editor} {\bibfnamefont {R.}~\bibnamefont {Lehnert}}}\
  (\bibinfo  {publisher} {{WORLD SCIENTIFIC}},\ \bibinfo {year} {2020})\ pp.\
  \bibinfo {pages} {37--40}\BibitemShut {NoStop}%
\bibitem [{\citenamefont {Pearle}(1989)}]{Pearle1989PRA}%
  \BibitemOpen
  \bibfield  {author} {\bibinfo {author} {\bibfnamefont {P.}~\bibnamefont
  {Pearle}},\ }\href {\doibase 10.1103/PhysRevA.39.2277} {\bibfield  {journal}
  {\bibinfo  {journal} {Phys. Rev. A}\ }\textbf {\bibinfo {volume} {39}},\
  \bibinfo {pages} {2277} (\bibinfo {year} {1989})}\BibitemShut {NoStop}%
\bibitem [{\citenamefont {Ghirardi}\ \emph {et~al.}(1990)\citenamefont
  {Ghirardi}, \citenamefont {Pearle},\ and\ \citenamefont
  {Rimini}}]{Ghirardi1990PRA}%
  \BibitemOpen
  \bibfield  {author} {\bibinfo {author} {\bibfnamefont {G.~C.}\ \bibnamefont
  {Ghirardi}}, \bibinfo {author} {\bibfnamefont {P.}~\bibnamefont {Pearle}}, \
  and\ \bibinfo {author} {\bibfnamefont {A.}~\bibnamefont {Rimini}},\ }\href
  {\doibase 10.1103/PhysRevA.42.78} {\bibfield  {journal} {\bibinfo  {journal}
  {Phys. Rev. A}\ }\textbf {\bibinfo {volume} {42}},\ \bibinfo {pages} {78}
  (\bibinfo {year} {1990})}\BibitemShut {NoStop}%
\bibitem [{\citenamefont {Carlesso}\ \emph
  {et~al.}(2022{\natexlab{b}})\citenamefont {Carlesso}, \citenamefont {Donadi},
  \citenamefont {Ferialdi}, \citenamefont {Paternostro}, \citenamefont
  {Ulbricht},\ and\ \citenamefont {Bassi}}]{Carlesso.2022}%
  \BibitemOpen
  \bibfield  {author} {\bibinfo {author} {\bibfnamefont {M.}~\bibnamefont
  {Carlesso}}, \bibinfo {author} {\bibfnamefont {S.}~\bibnamefont {Donadi}},
  \bibinfo {author} {\bibfnamefont {L.}~\bibnamefont {Ferialdi}}, \bibinfo
  {author} {\bibfnamefont {M.}~\bibnamefont {Paternostro}}, \bibinfo {author}
  {\bibfnamefont {H.}~\bibnamefont {Ulbricht}}, \ and\ \bibinfo {author}
  {\bibfnamefont {A.}~\bibnamefont {Bassi}},\ }\href {\doibase
  10.1038/s41567-021-01489-5} {\bibfield  {journal} {\bibinfo  {journal}
  {Nature Physics}\ }\textbf {\bibinfo {volume} {18}},\ \bibinfo {pages} {243}
  (\bibinfo {year} {2022}{\natexlab{b}})}\BibitemShut {NoStop}%
\bibitem [{\citenamefont {Roberts}\ \emph {et~al.}(1998)\citenamefont
  {Roberts}, \citenamefont {Claussen}, \citenamefont {Burke}, \citenamefont
  {Greene}, \citenamefont {Cornell},\ and\ \citenamefont
  {Wieman}}]{Roberts1998PRL}%
  \BibitemOpen
  \bibfield  {author} {\bibinfo {author} {\bibfnamefont {J.~L.}\ \bibnamefont
  {Roberts}}, \bibinfo {author} {\bibfnamefont {N.~R.}\ \bibnamefont
  {Claussen}}, \bibinfo {author} {\bibfnamefont {J.~P.}\ \bibnamefont {Burke}},
  \bibinfo {author} {\bibfnamefont {C.~H.}\ \bibnamefont {Greene}}, \bibinfo
  {author} {\bibfnamefont {E.~A.}\ \bibnamefont {Cornell}}, \ and\ \bibinfo
  {author} {\bibfnamefont {C.~E.}\ \bibnamefont {Wieman}},\ }\href {\doibase
  10.1103/PhysRevLett.81.5109} {\bibfield  {journal} {\bibinfo  {journal}
  {Phys. Rev. Lett.}\ }\textbf {\bibinfo {volume} {81}},\ \bibinfo {pages}
  {5109} (\bibinfo {year} {1998})}\BibitemShut {NoStop}%
\bibitem [{\citenamefont {Marte}\ \emph {et~al.}(2002)\citenamefont {Marte},
  \citenamefont {Volz}, \citenamefont {Schuster}, \citenamefont {D\"urr},
  \citenamefont {Rempe}, \citenamefont {van Kempen},\ and\ \citenamefont
  {Verhaar}}]{Marte2002PRL}%
  \BibitemOpen
  \bibfield  {author} {\bibinfo {author} {\bibfnamefont {A.}~\bibnamefont
  {Marte}}, \bibinfo {author} {\bibfnamefont {T.}~\bibnamefont {Volz}},
  \bibinfo {author} {\bibfnamefont {J.}~\bibnamefont {Schuster}}, \bibinfo
  {author} {\bibfnamefont {S.}~\bibnamefont {D\"urr}}, \bibinfo {author}
  {\bibfnamefont {G.}~\bibnamefont {Rempe}}, \bibinfo {author} {\bibfnamefont
  {E.~G.~M.}\ \bibnamefont {van Kempen}}, \ and\ \bibinfo {author}
  {\bibfnamefont {B.~J.}\ \bibnamefont {Verhaar}},\ }\href {\doibase
  10.1103/PhysRevLett.89.283202} {\bibfield  {journal} {\bibinfo  {journal}
  {Phys. Rev. Lett.}\ }\textbf {\bibinfo {volume} {89}},\ \bibinfo {pages}
  {283202} (\bibinfo {year} {2002})}\BibitemShut {NoStop}%
\bibitem [{\citenamefont {Koch}\ \emph {et~al.}(2005)\citenamefont {Koch},
  \citenamefont {Masnou-Seeuws},\ and\ \citenamefont {Kosloff}}]{Koch2005PRL}%
  \BibitemOpen
  \bibfield  {author} {\bibinfo {author} {\bibfnamefont {C.~P.}\ \bibnamefont
  {Koch}}, \bibinfo {author} {\bibfnamefont {F.~m.~c.}\ \bibnamefont
  {Masnou-Seeuws}}, \ and\ \bibinfo {author} {\bibfnamefont {R.}~\bibnamefont
  {Kosloff}},\ }\href {\doibase 10.1103/PhysRevLett.94.193001} {\bibfield
  {journal} {\bibinfo  {journal} {Phys. Rev. Lett.}\ }\textbf {\bibinfo
  {volume} {94}},\ \bibinfo {pages} {193001} (\bibinfo {year}
  {2005})}\BibitemShut {NoStop}%
\bibitem [{\citenamefont {Koch}(2008)}]{Koch2008PRA}%
  \BibitemOpen
  \bibfield  {author} {\bibinfo {author} {\bibfnamefont {C.~P.}\ \bibnamefont
  {Koch}},\ }\href {\doibase 10.1103/PhysRevA.78.063411} {\bibfield  {journal}
  {\bibinfo  {journal} {Phys. Rev. A}\ }\textbf {\bibinfo {volume} {78}},\
  \bibinfo {pages} {063411} (\bibinfo {year} {2008})}\BibitemShut {NoStop}%
\bibitem [{\citenamefont {Yan}\ \emph {et~al.}(2013)\citenamefont {Yan},
  \citenamefont {DeSalvo}, \citenamefont {Huang}, \citenamefont {Naidon},\ and\
  \citenamefont {Killian}}]{PhysRevLett.111.150402}%
  \BibitemOpen
  \bibfield  {author} {\bibinfo {author} {\bibfnamefont {M.}~\bibnamefont
  {Yan}}, \bibinfo {author} {\bibfnamefont {B.~J.}\ \bibnamefont {DeSalvo}},
  \bibinfo {author} {\bibfnamefont {Y.}~\bibnamefont {Huang}}, \bibinfo
  {author} {\bibfnamefont {P.}~\bibnamefont {Naidon}}, \ and\ \bibinfo {author}
  {\bibfnamefont {T.~C.}\ \bibnamefont {Killian}},\ }\href {\doibase
  10.1103/PhysRevLett.111.150402} {\bibfield  {journal} {\bibinfo  {journal}
  {Phys. Rev. Lett.}\ }\textbf {\bibinfo {volume} {111}},\ \bibinfo {pages}
  {150402} (\bibinfo {year} {2013})}\BibitemShut {NoStop}%
\bibitem [{\citenamefont {Taie}\ \emph {et~al.}(2016)\citenamefont {Taie},
  \citenamefont {Watanabe}, \citenamefont {Ichinose},\ and\ \citenamefont
  {Takahashi}}]{PhysRevLett.116.043202}%
  \BibitemOpen
  \bibfield  {author} {\bibinfo {author} {\bibfnamefont {S.}~\bibnamefont
  {Taie}}, \bibinfo {author} {\bibfnamefont {S.}~\bibnamefont {Watanabe}},
  \bibinfo {author} {\bibfnamefont {T.}~\bibnamefont {Ichinose}}, \ and\
  \bibinfo {author} {\bibfnamefont {Y.}~\bibnamefont {Takahashi}},\ }\href
  {\doibase 10.1103/PhysRevLett.116.043202} {\bibfield  {journal} {\bibinfo
  {journal} {Phys. Rev. Lett.}\ }\textbf {\bibinfo {volume} {116}},\ \bibinfo
  {pages} {043202} (\bibinfo {year} {2016})}\BibitemShut {NoStop}%
\bibitem [{\citenamefont {Theis}\ \emph {et~al.}(2004)\citenamefont {Theis},
  \citenamefont {Thalhammer}, \citenamefont {Winkler}, \citenamefont {Hellwig},
  \citenamefont {Ruff}, \citenamefont {Grimm},\ and\ \citenamefont
  {Denschlag}}]{Theis2004PRL}%
  \BibitemOpen
  \bibfield  {author} {\bibinfo {author} {\bibfnamefont {M.}~\bibnamefont
  {Theis}}, \bibinfo {author} {\bibfnamefont {G.}~\bibnamefont {Thalhammer}},
  \bibinfo {author} {\bibfnamefont {K.}~\bibnamefont {Winkler}}, \bibinfo
  {author} {\bibfnamefont {M.}~\bibnamefont {Hellwig}}, \bibinfo {author}
  {\bibfnamefont {G.}~\bibnamefont {Ruff}}, \bibinfo {author} {\bibfnamefont
  {R.}~\bibnamefont {Grimm}}, \ and\ \bibinfo {author} {\bibfnamefont {J.~H.}\
  \bibnamefont {Denschlag}},\ }\href {\doibase 10.1103/PhysRevLett.93.123001}
  {\bibfield  {journal} {\bibinfo  {journal} {Phys. Rev. Lett.}\ }\textbf
  {\bibinfo {volume} {93}},\ \bibinfo {pages} {123001} (\bibinfo {year}
  {2004})}\BibitemShut {NoStop}%
\bibitem [{\citenamefont {Ciury\l{}o}\ \emph {et~al.}(2005)\citenamefont
  {Ciury\l{}o}, \citenamefont {Tiesinga},\ and\ \citenamefont
  {Julienne}}]{Ciuryo2005PRL}%
  \BibitemOpen
  \bibfield  {author} {\bibinfo {author} {\bibfnamefont {R.}~\bibnamefont
  {Ciury\l{}o}}, \bibinfo {author} {\bibfnamefont {E.}~\bibnamefont
  {Tiesinga}}, \ and\ \bibinfo {author} {\bibfnamefont {P.~S.}\ \bibnamefont
  {Julienne}},\ }\href {\doibase 10.1103/PhysRevA.71.030701} {\bibfield
  {journal} {\bibinfo  {journal} {Phys. Rev. A}\ }\textbf {\bibinfo {volume}
  {71}},\ \bibinfo {pages} {030701} (\bibinfo {year} {2005})}\BibitemShut
  {NoStop}%
\bibitem [{\citenamefont {Enomoto}\ \emph {et~al.}(2008)\citenamefont
  {Enomoto}, \citenamefont {Kasa}, \citenamefont {Kitagawa},\ and\
  \citenamefont {Takahashi}}]{Enomoto2008PRL}%
  \BibitemOpen
  \bibfield  {author} {\bibinfo {author} {\bibfnamefont {K.}~\bibnamefont
  {Enomoto}}, \bibinfo {author} {\bibfnamefont {K.}~\bibnamefont {Kasa}},
  \bibinfo {author} {\bibfnamefont {M.}~\bibnamefont {Kitagawa}}, \ and\
  \bibinfo {author} {\bibfnamefont {Y.}~\bibnamefont {Takahashi}},\ }\href
  {\doibase 10.1103/PhysRevLett.101.203201} {\bibfield  {journal} {\bibinfo
  {journal} {Phys. Rev. Lett.}\ }\textbf {\bibinfo {volume} {101}},\ \bibinfo
  {pages} {203201} (\bibinfo {year} {2008})}\BibitemShut {NoStop}%
\bibitem [{\citenamefont {Blatt}\ \emph {et~al.}(2011)\citenamefont {Blatt},
  \citenamefont {Nicholson}, \citenamefont {Bloom}, \citenamefont {Williams},
  \citenamefont {Thomsen}, \citenamefont {Julienne},\ and\ \citenamefont
  {Ye}}]{Blatt11PRL}%
  \BibitemOpen
  \bibfield  {author} {\bibinfo {author} {\bibfnamefont {S.}~\bibnamefont
  {Blatt}}, \bibinfo {author} {\bibfnamefont {T.~L.}\ \bibnamefont
  {Nicholson}}, \bibinfo {author} {\bibfnamefont {B.~J.}\ \bibnamefont
  {Bloom}}, \bibinfo {author} {\bibfnamefont {J.~R.}\ \bibnamefont {Williams}},
  \bibinfo {author} {\bibfnamefont {J.~W.}\ \bibnamefont {Thomsen}}, \bibinfo
  {author} {\bibfnamefont {P.~S.}\ \bibnamefont {Julienne}}, \ and\ \bibinfo
  {author} {\bibfnamefont {J.}~\bibnamefont {Ye}},\ }\href {\doibase
  10.1103/PhysRevLett.107.073202} {\bibfield  {journal} {\bibinfo  {journal}
  {Phys. Rev. Lett.}\ }\textbf {\bibinfo {volume} {107}},\ \bibinfo {pages}
  {073202} (\bibinfo {year} {2011})}\BibitemShut {NoStop}%
\bibitem [{\citenamefont {Zelevinsky}\ \emph {et~al.}(2006)\citenamefont
  {Zelevinsky}, \citenamefont {Boyd}, \citenamefont {Ludlow}, \citenamefont
  {Ido}, \citenamefont {Ye}, \citenamefont {Ciury\l{}o}, \citenamefont
  {Naidon},\ and\ \citenamefont {Julienne}}]{Zelevinsky2006PRL}%
  \BibitemOpen
  \bibfield  {author} {\bibinfo {author} {\bibfnamefont {T.}~\bibnamefont
  {Zelevinsky}}, \bibinfo {author} {\bibfnamefont {M.~M.}\ \bibnamefont
  {Boyd}}, \bibinfo {author} {\bibfnamefont {A.~D.}\ \bibnamefont {Ludlow}},
  \bibinfo {author} {\bibfnamefont {T.}~\bibnamefont {Ido}}, \bibinfo {author}
  {\bibfnamefont {J.}~\bibnamefont {Ye}}, \bibinfo {author} {\bibfnamefont
  {R.}~\bibnamefont {Ciury\l{}o}}, \bibinfo {author} {\bibfnamefont
  {P.}~\bibnamefont {Naidon}}, \ and\ \bibinfo {author} {\bibfnamefont {P.~S.}\
  \bibnamefont {Julienne}},\ }\href {\doibase 10.1103/PhysRevLett.96.203201}
  {\bibfield  {journal} {\bibinfo  {journal} {Phys. Rev. Lett.}\ }\textbf
  {\bibinfo {volume} {96}},\ \bibinfo {pages} {203201} (\bibinfo {year}
  {2006})}\BibitemShut {NoStop}%
\end{thebibliography}%
\end{document}